\documentclass[%
amsmath,amssymb,%
superscriptaddress,
reprint,%
aps,dvipsnames,notitlepage
]{revtex4-2}
\usepackage{upgreek}
\usepackage{booktabs}
\usepackage{siunitx}

\usepackage{graphicx} 
\usepackage{dcolumn} 
\usepackage{bm} 
\usepackage{braket}
\usepackage{xcolor}
\definecolor{edcolor}{RGB}{100, 180, 240}
\definecolor{jscolor}{RGB}{178, 110, 181}

\usepackage{dsfont} 
\usepackage{soul}

\usepackage{bbold}

\usepackage{hyperref}
\usepackage{enumitem}
\usepackage{anyfontsize,amsfonts} 

\usepackage[mathlines]{lineno} 
\modulolinenumbers[5] 

\newcommand{\eju}{E_{\mathrm{J}_1}}
\newcommand{\ejd}{E_{\mathrm{J}_2}}
\newcommand{\ec}{E_\mathrm{C}}
\newcommand{\ejsu}{E_{\mathrm{J}\Sigma_1}}
\newcommand{\ejsd}{E_{\mathrm{J}\Sigma_2}}

\usepackage{array,tabularx,csquotes}
\renewcommand{\dag}{^{\dagger}}
\newcommand{\half}{\frac{1}{2}}
\newcommand{\rate}[2]{\Gamma_{#1 \rightarrow #2}}
\newcommand{\dm}{\hat{\rho}}
\newcommand{\phiop}{\widehat{\varphi}}
\newcommand{\ketbra}[2]{\ket{#1}\!\!\bra{#2}}

\AtBeginEnvironment{pmatrix}{\everymath{\displaystyle}}
\hypersetup{colorlinks,linkcolor={red!50!black},citecolor={blue!50!black},urlcolor={blue!80!black}}

\begin{document}




\title{Coherence Limits in Interference-Based cos(2\texorpdfstring{\bm{$\varphi$}}{phi}) Qubits}

\author{S. Messelot}
\thanks{Present address: Quobly, 38000 Grenoble, France}
\affiliation{Univ. Grenoble Alpes, CNRS, Grenoble INP, Institut Néel, 38000 Grenoble, France}
\author{A. Leblanc}
\thanks{Present address: Center for Quantum Information Physics, Department of Physics, New York University, New York, NY 10003, USA}
\affiliation{Univ. Grenoble Alpes, CEA, Grenoble INP, IRIG, PHELIQS, 38000 Grenoble, France}
\author{J.-S. Tettekpoe}
\affiliation{Univ. Grenoble Alpes, CNRS, Grenoble INP, Institut Néel, 38000 Grenoble, France}
\author{F. Lefloch}
\affiliation{Univ. Grenoble Alpes, CEA, Grenoble INP, IRIG, PHELIQS, 38000 Grenoble, France}
\author{Q. Ficheux}
\email{quentin.ficheux@neel.cnrs.fr}
\affiliation{Univ. Grenoble Alpes, CNRS, Grenoble INP, Institut Néel, 38000 Grenoble, France}
\author{J. Renard}
\affiliation{Univ. Grenoble Alpes, CNRS, Grenoble INP, Institut Néel, 38000 Grenoble, France}

\author{É. Dumur}
\email{etienne.dumur@cea.fr}
\affiliation{Univ. Grenoble Alpes, CEA, Grenoble INP, IRIG, PHELIQS, 38000 Grenoble, France}

\date{\today}

\begin{abstract}

We investigate the coherence properties of parity-protected $\cos(2\varphi)$ qubits based on interferences between two Josephson elements in a superconducting loop. We show that qubit implementations of a $\cos(2\varphi)$ potential using a single loop, such as those employing semiconducting junctions, rhombus circuits, flowermon and KITE structures, can be described by the same Hamiltonian as two multi-harmonic Josephson junctions in a SQUID geometry. We find that, despite the parity protection arising from the suppression of single Cooper pair tunneling, there exists a fundamental trade-off between charge and flux noise dephasing channels. Using numerical simulations, we examine how relaxation and dephasing rates depend on external flux and circuit parameters, and we identify the best compromise for maximum coherence. With currently existing circuit parameters, the qubit lifetime $T_1$ can exceed milliseconds while the dephasing time $T_\varphi$ remains limited to only a few microseconds due to either flux or charge noise.
Our findings establish practical limits on the coherence of this class of qubits and raise questions about the long-term potential of this approach.

\end{abstract}

\maketitle

The transmon qubit~\cite{Koch2007} is arguably the most widely used superconducting qubit due to its simplicity and long coherence times, with state-of-the-art decay and dephasing times reaching $T_1, T_\varphi \sim$ \SI{1}{\milli\second}~\cite{Tuokkola2024,bland20252d}. Interestingly, the transmon qubit was originally introduced as a ``charge insensitive qubit'' extending the charge noise protection of the quantronium~\cite{vion2002manipulating} from first to all orders. As the ratio of Josephson energy to charging energy, $E_\mathrm{J}/\ec$, increases, the sensitivity to charge noise decreases exponentially, providing excellent protection against dephasing. However, the qubit remains susceptible to energy decay via electrical coupling to environmental degrees of freedom, owing to its large charge matrix element of the order of unity.


Noise protection against dephasing can be extended to create qubits that are also resistant to energy relaxation~\cite{Blatter2001, Doucot2002, Protopopov2004, gladchenko2009superconducting, Bell2014}.
The key concept behind this protection is to leverage the internal symmetry of the system, encoding the ground and excited states with wave functions that have different parities~\cite{gyenis2021moving,gyenis2019experimental,Somoroff2023,Earnest2018,wang2024quantum,kalashnikov2020bifluxon}.
For example, the Hamiltonian of the circuit can be engineered to preserve the parity of the number of Cooper pairs by only allowing the coherent tunneling of pairs of Cooper pairs through an operator $|N\rangle \langle N+2 | + |N+2\rangle \langle N|$, which couples the charge states of the same parity.
This is known as the $\cos(2\varphi)$ qubit.
In practice, it resembles a transmon, but instead of a Josephson junction with a $2\pi$-periodic potential $-E_{\mathrm{J}_1} \cos(\varphi) $, we have a $\pi$-periodic circuit element.
This element allows only pairs of Cooper pairs to be transmitted, resulting in a potential described by $E_{\mathrm{J}_2} \cos(2\varphi)$.
The ground and excited state wavefunctions have different parities in the charge basis, leading to disjoint wavefunction supports and, therefore, suppressed matrix elements.





One method to engineer a $\pi$-periodic potential is to leverage phase interferences between two circuit elements with non-sinusoidal current-phase relations embedded in a loop. When the loop is biased at half a magnetic flux quantum, the odd harmonics of the current interfere destructively, while the even harmonics remain.

Several interference-based parity-protected circuits have been proposed, such as the Rhombus~\cite{Doucot2002}, the Bifluxon~\cite{kalashnikov2020bifluxon}, the Kinetic Interference coTunneling Element (KITE)~\cite{Smith2020, alvise} and the 0-$\pi$~\cite{Paolo2019}. The aforementioned circuits rely on standard aluminum - aluminum oxide (Al/AlOx) Josephson junctions, and current-phase relation (CPR) higher harmonics stem from the detailed architecture of the $\cos(2\varphi)$ element. Alternatively, intrinsically multi-harmonic Josephson junctions based on semiconductor weak links have been demonstrated, using material platforms such as InAs nanowires~\cite{spanton2017current}, InAs~\cite{zhang2024large,Liu2025} and Ge/SiGe~\cite{Leblanc2024} two dimensional electron gas, graphene~\cite{nanda2017current, Messelot2024a} and other 2D materials~\cite{endres2023current}.
Other intrinsically multi-harmonic junctions include pinhole Al junctions~\cite{Griesmar2025} or twisted cuprate van der Waals heterostructures~\cite{flowermon}.
However, only a handful of experiments~\cite{Bell2014,Larsen2020} have successfully demonstrated effective protection against energy relaxation using $\cos(2\varphi)$ elements.
To date, none have managed to combine this protection with protection against dephasing.
This naturally raises the question of potential decoherence mechanisms specific to such an implementation.\\

In this article, we conduct a comprehensive investigation of coherence properties of a general model of $\cos(2\varphi)$ qubits, where Cooper-pair pairing is achieved through interferences between two bi-harmonic Josephson elements.
We study the resilience of the energy relaxation protection as a function of circuit parameters, imperfections, and finite temperature effects.
Furthermore, we show that the introduction of a loop in the circuit brings an unavoidable trade-off between charge and flux noise protection for the dephasing time.
We show that biasing the circuit loop away from the frustration point can be advantageous, as it helps to balance the circuit's susceptibility to these two noise sources and allows to reach coherence times of several microseconds with currently experimentally accessible circuit parameters.

We start with a detailed description of the circuit, its Hamiltonian, and the various platforms that implement it. This is followed by a discussion of the effects of magnetic flux and charge offset. In the second part, we focus on energy relaxation protection mechanisms and their robustness to variations in circuit parameters. In a third part, we describe the specific coherence issue faced by the $\cos\left( 2 \varphi\right)$ qubit, and in the fourth part, we optimize the circuit parameters to achieve maximum coherence.

\section{Interference-based cos(2\texorpdfstring{\bm{$\varphi$}}{phi})}
\subsection{Hamiltonian}

\begin{figure}[ht]
    \centering
    \includegraphics[width=\linewidth]{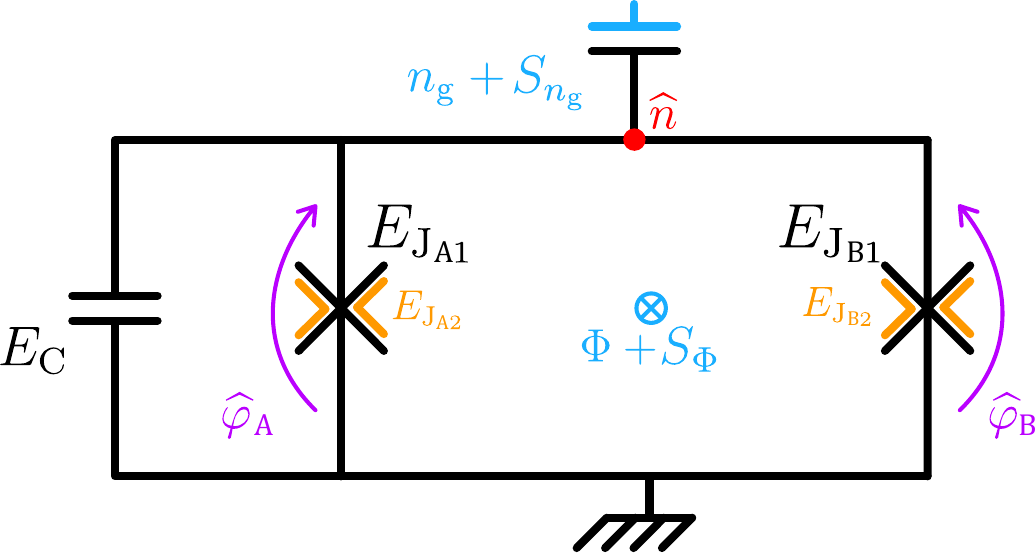}
    \caption{\textbf{Electrical circuit.}
    The circuit is made of two junctions A and B in parallel. Each junction has its own Josephson energies ($E_\mathrm{J_{i j}}$), where $i=A,B$ denotes the junction and $j=1,2$ is the order of the harmonic of the Josephson junction. A large shunt capacitance creates the charging energy $\ec$. The circuit has three internal quantum operators: the charge number of the common metallic island $\widehat{n}$, in red, and the two phases of the junctions $\widehat{\varphi}_{A,B}$, in purple. It can be controlled via two classical external knobs shown in blue with the applied voltage, $n_\mathrm{g}$, and the applied magnetic flux, $\Phi$. The same control knobs couple to external noise densities denoted $S_{n_\mathrm{g}}$ and $S_\Phi$, electrical and magnetic, respectively.}
    \label{fig:circuit}
\end{figure}

The circuit investigated in this paper is composed of two bi-harmonic Josephson junctions, assembled in a superconducting quantum interference device (SQUID) configuration and shunted by a single capacitance. It is shown in Fig.~\ref{fig:circuit}.
We use as definition of the junction's Josephson potential $\widehat{U} = \eju \cos(\widehat{\varphi}) + \ejd \cos(2\widehat{\varphi})$~\cite{Hays2025}. With this definition, in typical multi-harmonic junctions, $\eju < 0$ and $\ejd>0$.
From standard circuit quantum electrodynamics (cQED) formalism~\cite{Devoret1997}, we find the Hamiltonian
\begin{equation}
    \begin{aligned}
        \widehat{H} &= \; 4 \ec \left(\widehat{n}  - n_g\right)^2 \\
        & \;\;\;+ E_\mathrm{J_{A1}} \cos\left(\widehat{\varphi}_A\right) + E_\mathrm{J_{B1}} \cos\left(\widehat{\varphi}_B  \right) \\
              & \;\;\;+ E_\mathrm{J_{A2}} \cos\left(2\widehat{\varphi}_A\right)  + E_\mathrm{J_{B2}} \cos\left(2\widehat{\varphi}_B\right)
        \label{eq:ham1}
    \end{aligned}
\end{equation}
with the electron charging energy $\ec= e^2/(2 C)$, the Josephson energy $E_\mathrm{J_{ij}} = \Phi_0 I_\mathrm{c_{i, j}} /(2 \pi \times j) $ of junction $i$ and $j$-th harmonics,
$\widehat{n}$ is the quantum charge operator of the superconducting island, and $n_\mathrm{g}$ the classical charge offset. Similarly, $\widehat{\varphi}_i$ is the quantum phase operator of junction $i$.
Compared to a ``pure'' $\cos(2\varphi)$ Hamiltonian $\widehat{H} = 4 \ec \left(\widehat{n}  - n_g\right)^2 + \ejd\cos\left(2\widehat{\varphi}\right)$, the ``interference-based'' $\cos(2\varphi)$ Hamiltonian encompasses the first harmonic terms of each junction that come with any practical realization.
The quantum phase operator $\widehat{\varphi}_i$ of the two junctions obeys the flux quantization of the superconducting loop:
\begin{equation}
    \varphi_A - \varphi_B = 2 \pi n + 2 \pi \frac{\Phi}{\Phi_0}, \quad \mathrm{with}\ n \in \mathbb{Z}
    \label{eq:flux-quantization}
\end{equation}
where $\Phi$ is the classical flux offset that can be externally controlled and ${\Phi_0}$ the magnetic flux quantum.
We choose $\widehat{\varphi}=(\widehat{\varphi}_A+\widehat{\varphi}_B)/2$ as gauge for the phase degree of freedom, which satisfies the irrotational constraint~\cite{you2019circuit} provided the two junctions have equal self capacitances (that we omitted in Fig.~\ref{fig:circuit}, for clarity). This condition is verified in the following as we consider a close to symmetric SQUID, which allows direct calculation of coherence times~\cite{you2019circuit}.
Using Eq.~\eqref{eq:flux-quantization}, we can write the Hamiltonian of Eq.~\eqref{eq:ham1} as
\begin{equation}
    \begin{aligned}
        \widehat{H} & = \; 4 \ec \left(\widehat{n}  - n_g\right)^2 & \\
        &  \;\;\;+ \ejsu \left[  \; \; \; \; \;\cos\left(\phantom{2}\widehat{\varphi}\right) \cos\left(\phantom{2}\pi\left( \delta \Phi + \frac{1}{2}\right) \right) \right. & \\
        & \;\;\;\;\;\;\;\;\;\;\;\;\;\;\; \;  \left. + d_1 \sin\left(\phantom{2}\widehat{\varphi}  \right) \sin\left(\phantom{2}\pi\left( \delta \Phi + \frac{1}{2}\right) \right)\right] & \\
        & \;\;\; + \ejsd \left[  \; \; \; \; \;\cos\left(2\widehat{\varphi}\right) \cos\left(2\pi\left( \delta \Phi + \frac{1}{2}\right) \right) \right. & \\
        & \;\;\;\;\;\;\;\;\;\;\;\;\;\;\; \;  \left. + d_2 \sin\left(2\widehat{\varphi}  \right) \sin\left(2\pi\left( \delta \Phi + \frac{1}{2}\right) \right) \right]
    \label{eq:ham2}
    \end{aligned}
\end{equation}
with the SQUID Josephson energy $E_{\mathrm{J}\Sigma_i} = E_{\mathrm{J}_{Ai}} + E_{\mathrm{J}_{Bi}}$ and the junction asymmetries $d_i = (E_{\mathrm{J}_{Bi}} - E_{\mathrm{J}_{Ai}})/(E_{\mathrm{J}_{Bi}} + E_{\mathrm{J}_{Ai}})$.
As we  will focus on variations of flux close to $\Phi = \Phi_0 /2$ in the following discussion, we also introduce $\delta \Phi = \Phi/\Phi_0 - 1/2$, i.e. the offset flux with respect to the frustration point of the SQUID.
We note that with the first two terms of Eq.~\eqref{eq:ham2}, we recover the ``split'' transmon~\cite{Koch2007} while the last term is solely due to the second harmonic of the junctions. For simplicity, we will consider $d_1 = d_2 = d$ in the following, which amounts to assuming qualitatively similar Josephson elements A and B. In practice, only the first Josephson harmonics need precise symmetry to cancel out.

There are 6 circuit parameters, either set by design or controlled in-situ: $\ejsd/\ejsu$, $\ejsd/\ec$, $\ejsd \times \ec$, $\delta \Phi$, $n_g$, $d$. We will explore the effect of each of these parameters on the coherence properties of the $\cos(2\varphi)$ qubit calculated using the scQubits~\cite{Chitta_2022} and QuTiP~\cite{lambert2024qutip5quantumtoolbox} Python packages.

\subsection{Implementations}

\begin{table}[ht]
    \centering
    \begin{tabular}{ll}
\toprule
 Circuit & $\ejsd/\ejsu$\\
\midrule
Superconducting circuit based & \\
  $\;\;\;\;$ $\phantom{^{*,\,,\,}}$Rhombus~\cite{Doucot2002,Bell2014} & -0.2 \\
  $\;\;\;\;$ $\phantom{^{*,\,,\,}}$Pinhole JJ~\cite{Griesmar2025} & -0.1 \\
  $\;\;\;\;$ $\phantom{^{*,\,,\,}}$KITE (Low inductance)~\cite{alvise} & -0.025\\
  $\;\;\;\;$ $\phantom{^{*,\,}}^*$KITE (High inductance)~\cite{Smith2020} & -0.04 \\
Semiconductor  & \\
  $\;\;\;\;$ $\phantom{^{*,\,\ddagger}}$Germanium~\cite{Leblanc2024} & -0.1 \\
  $\;\;\;\;$ $\phantom{^{*,\,\ddagger}}$Graphene~\cite{Messelot2024a} & -0.1 \\
  $\;\;\;\;$ $\phantom{^{*,\,\ddagger}}$InAs~\cite{spanton2017current} & -0.1 -- -0.2 \\
Heterostructures & \\
  $\;\;\;\;$ $^{\dagger,\,\ddagger}$Flowermon~\cite{flowermon, biflowermon} & ±0.1 -- ±15\\
\bottomrule
    \end{tabular}
    \caption{\textbf{Interference-based cos(2\texorpdfstring{\bm{$\varphi$}}{phi}) implementations.}
    $^*$, estimated by fitting the energy spectrum from \cite{Smith2020} to the Hamiltonian given by Eq.~(\ref{eq:ham2}).
    $^\dagger$, only theoretical proposal at the time of this report.
    $^\ddagger$, the sign of $\ejsu$ may be positive or negative as opposed to other implementations.
    }
    \label{tab:all-circuits}
\end{table}

The circuit in Fig.~\ref{fig:circuit} and its corresponding Hamiltonian in Eq.~\eqref{eq:ham2}, while simple, encompass numerous experimental attempts to implement a $\cos(2\varphi)$ qubit, see Table ~\ref{tab:all-circuits}.
We can see that the means used to implement a $\cos(2\varphi)$ are diverse, including  superconducting circuits but also semi-conducting Josephson junctions and even heterostructures.

For each of the proposed implementations, we reported the  $\ejsd/\ejsu$ ratio, see details in Appendix~\ref{app:A}.
It is interesting to note that, while very different in nature, all current efforts show a similar $\ejsd/\ejsu$ ratio of $-0.1$, motivating the use of this value for the remainder of the paper.
An exception to this, that we discuss at the end of the paper is the Flowermon~\cite{flowermon, biflowermon} which, if successfully implemented, could reach a ratio on the order of -15.

\subsection{Interferences}

\begin{figure}[ht]
    \centering
    \includegraphics[width=\linewidth]{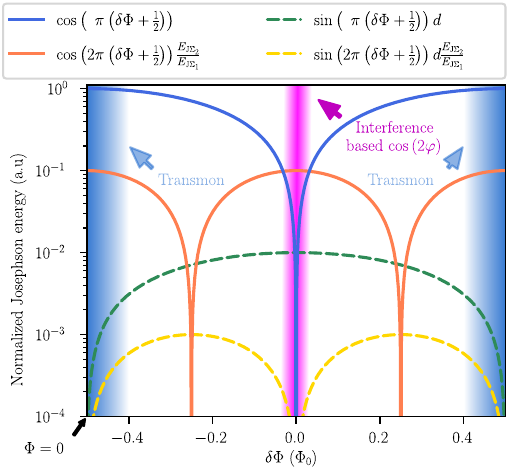}
    \caption{\textbf{Magnetic flux dependence on the junctions harmonics energy.}
    Magnetic flux dependence of the Josephson harmonics, see Eq.~\eqref{eq:ham2}.
    With a small junctions asymmetry, two regimes can be reached at $\delta \Phi \sim \pm 1/2$ ($\Phi / \Phi_0 \sim 0$), where the physics is similar to a transmon, and at $\delta \Phi \sim 0$, where the $\cos(2\varphi)$ is dominant.
    We took $\ejsd/\ejsu=-0.1$ and $d = \SI{1}{\percent}$.
    }
    \label{fig:trigo}
\end{figure}

In Fig.~\ref{fig:trigo}, we show the magnetic flux dependence of the Josephson terms in Eq.~\eqref{eq:ham2}.
We see that as a function of the magnetic flux, the dominant Josephson harmonics, in blue and orange solid lines for the 1\textsuperscript{st} and 2\textsuperscript{nd} ones, are swapped.
As a direct consequence, two regimes can be considered: near $\delta \Phi \sim \pm 0.5$ ($\Phi/\Phi_0 \sim 0$) the circuit is similar to a transmon, and near $\delta \Phi \sim 0$, the $\cos(2\varphi)$ behavior is the relevant one.
Due to inevitable junction asymmetry, out-of-phase supplemental terms arise, shown as dashed green and gold lines for the 1\textsuperscript{st} and 2\textsuperscript{nd} harmonics. The first and most important effect is that the 1\textsuperscript{st} harmonic does not exactly cancel out at $\delta \Phi =0$.
Hence, it is of crucial importance to have highly symmetric junctions
to keep the 2\textsuperscript{nd} harmonic dominant in this regime.
In the case of hybrid semiconductor Josephson junctions, the Josephson energy can be tuned \emph{in-situ} allowing asymmetry $d \lesssim \SI{1}{\percent}$ to be achieved routinely~\cite{Leblanc2025}, while exhibiting a significant second order Josephson energy $\ejsd$.
Standard aluminum Josephson junctions reach asymmetry $d \approx \SI{1}{\percent}$ on wafer scale~\cite{Kreikebaum2020} making them also suitable.
This makes the interference-based $\cos(2\varphi)$ regime experimentally accessible in these platforms.



\subsection{Wave functions}

\begin{figure}[ht]
    \centering
    \includegraphics[width=\linewidth]{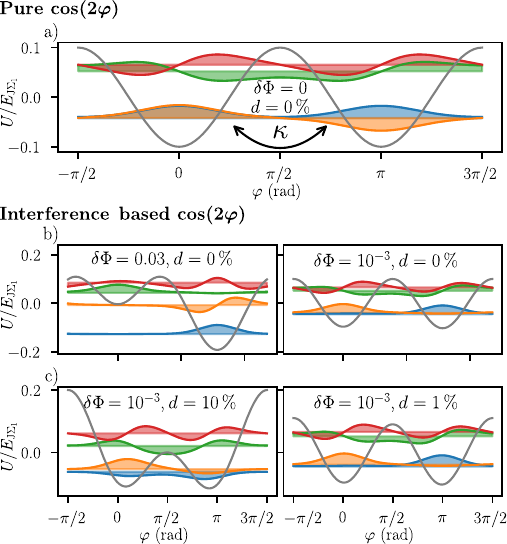}
    \caption{\textbf{Impact of magnetic flux offset and junctions asymmetry on the wavefunctions}.
    a) Pure $\cos(2\varphi)$. The lowest eigenstates are delocalized in the two wells
    due to the inter-well interaction energy $\kappa$.
    b) Influence of the offset magnetic flux $\delta \Phi$ with asymmetry set to zero.
    c) Influence of the junctions asymmetry with the offset magnetic flux $\delta \Phi = 10^{-3}$.
    For all figures, we have $\ejsd/\ejsu=-0.1$ and $\ejsd/\ec = 20$.}
    \label{fig:potential}
\end{figure}

The interest of the $\cos(2\varphi)$ qubit is rooted in its expected long lifetime~\cite{Doucot2002,Larsen2020, Smith2020,Hays2025}. The tunneling of single Cooper pairs is in principle forbidden in such a circuit, ensuring that the transition matrix element $\langle e | \widehat{n} | g \rangle$ between ground $| g \rangle$ and excited states $| e \rangle$ vanishes. Only the tunneling of pairs of Cooper pairs is allowed.
However, due to junction asymmetry or imperfect magnetic flux parking, the Hamiltonian has a residual
$\cos(1\varphi)$ term, see Fig.~\ref{fig:trigo}.
In the following, we will discuss
protection against relaxation in these conditions.

One way to address this question is to look at the eigenstates' wavefunctions represented in Fig.~\ref{fig:potential}.
A pure $\cos(2\varphi)$ system exhibits a characteristic two-well potential. The two lowest-lying states are symmetric and anti-symmetric superposition of states localized in each well.
The amplitude of the matrix element $\langle0|\widehat{n}|1\rangle$ is exactly 0.
Considering now an interference-based $\cos(2\varphi)$ qubit, if $\delta \Phi$ as well as $d$ are precisely equal to 0, the eigenfunctions are identical to that of a pure $\cos(2\varphi)$ system.

Non-zero flux offset $\delta \Phi$ and asymmetry $d$ result in significant modification of this picture. In Fig.~\ref{fig:potential}.b) left panel, at $\delta \Phi = 0.03$, the circuit is close to the $\cos(2\varphi)$ regime as we can observe two potential wells.
However, the first two wavefunctions are in the same well due to the remaining $\cos(\varphi)$ term and this regime offers no additional protection compared to the transmon.
Now in the right-panel, we set a smaller $\delta \Phi = 10^{-3}$, which is compatible with experimentally achievable flux resolution~\cite{Quintana2017}.
The $\cos(\varphi)$ term is sufficiently suppressed and consequently, both wavefunctions are localized in different wells. These two states can be described by opposite persistent current in the SQUID loop.
They remain however non-degenerate since the offset flux creates an imbalance between the two potential wells greater than the inter-well coupling energy $\kappa$.

In Fig.~\ref{fig:potential}.c), the magnetic flux is kept close to half a flux quanta $\delta \Phi = 10^{-3}$ and the junction asymmetry $d$ is  set to \SI{10}{\percent} (left panel) and \SI{1}{\percent} (right panel).
We observe that a \SI{10}{\percent} asymmetry results in a substantial overlap between the ground and first excited state wavefunctions within the same potential well.
This effectively breaks the protection from energy relaxation.
However at a smaller asymmetry of \SI{1}{\percent}, an experimentally reachable parameter, both wavefunctions stay localized in different wells.

The conclusions are threefold. First despite the fact that they both re-introduce a $\cos(1\varphi)$ term in the Josephson potential, offset flux and asymmetry have different effects: offset flux lifts the degeneracy between the two wells whereas junction asymmetry weakens the potential barrier between them.
Second, since the degeneracy lifting at $\delta \Phi = 0$ results from inter-well coupling through the phase potential barrier, the energy splitting is exponentially suppressed with $\ejsd / \ec$, potentially resulting in MHz or even lower qubit transitions.
Finally, while the circuit is not in a pure $\cos( 2 \varphi)$ regime due to finite precision in the junctions' symmetry and the magnetic flux, it can still localize the two wavefunctions in different wells. The consequence is a long lifetime as the matrix element $\langle e | \widehat{n} | g \rangle = - i \langle e | \partial/\partial \widehat{\varphi} | g \rangle $ is exponentially suppressed with $\ejsd/\ec$.



\subsection{Charge dispersion}

\begin{figure}[ht]
    \centering
    \includegraphics[width=\linewidth]{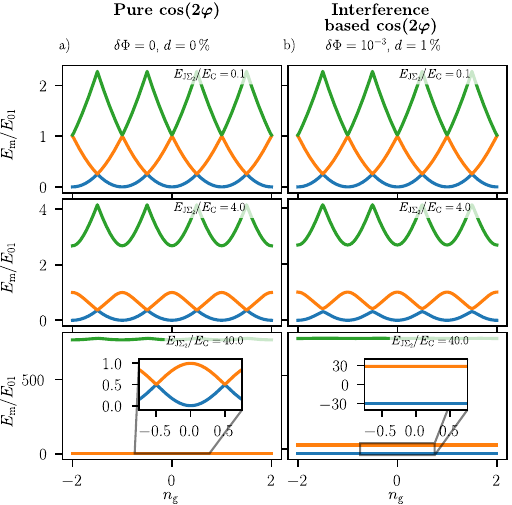}
    \caption{Eigenenergies $E_m$, $ m \in \{0,1,2\}$, of the circuit Hamiltonian, see Eq.~\eqref{eq:ham2}, as a function of the offset charge $n_g$ for different ratios $E_{\mathrm{J}\Sigma_{\{1,2\}}}/\ec$, linewise, and offset flux $\Phi$, columnwise.
    Energies are given in units of the transition energy $E_{01}$ of the pure $\cos (2 \varphi)$, evaluated at  $n_g=0$.
    The zero point of energy is chosen as the bottom of the $m=0$ level.
    The left column shows the pure $\cos (2 \varphi)$ regime, at exactly $\delta \Phi = 0$ where only the second harmonics are present.
    The right column shows the interference-based $\cos (2 \varphi)$ case, where $\delta \Phi = 10^{-3}$.
    For all figures, we have $\ejsd/\ejsu=-0.1$.
    }
    \label{fig:transmon-vs-cos2phi}
\end{figure}

In Fig.~\ref{fig:transmon-vs-cos2phi}, we show the three lowest eigenenergies as function of the island offset charge $n_\mathrm{g}$ for several values of the $E_{\mathrm{J}\Sigma_{\{1,2\}}}/\ec$ ratio, increasing from top to bottom.
In Fig.~\ref{fig:transmon-vs-cos2phi}.a), at exactly $\delta \Phi = 0$ and with $d=\SI{0}{\percent}$, we explore the pure $\cos(2\varphi)$ regime, which, despite being experimentally unachievable, has striking features.
Its charge dispersion is equal to the transition frequency $f_{01}$ at $n_\mathrm{g}=0$, which prevents the charge dispersion from flattening.
Indeed,
since $n_\mathrm{g} =0.5$ is a symmetry point between even and odd charge states, the two lowest-lying states have exactly the same energy for this particular value of $n_\mathrm{g}$.
This degeneracy is not lifted because of the absence of single Cooper pair tunneling in the pure $\cos(2\varphi)$ regime.

In Fig.~\ref{fig:transmon-vs-cos2phi}.b), at $\delta \Phi = 10^{-3}$ and with $d=\SI{1}{\percent}$, we now explore the interference-based $\cos(2\varphi)$ case.
The most noticeable difference with the pure regime is the recovery of a flat charge dispersion at large $\ejsd/\ec$ ratio due to the reintroduction of a small $\cos(1\varphi)$ component.
We see that at large $\ejsd/\ec$ ratio, the third energy level stays extremely far from the second one, indicating a very large qubit anharmonicity, which is lacking in the transmon.
This effect is due to the localization of the wavefunctions in the left or right well of the two-well potential as seen in Fig.~\ref{fig:potential}.
Indeed, there are two types of transitions in the $\cos(2\varphi)$ regime: intra-well transitions (plasmons), here from ground state to 2\textsuperscript{nd} excited state, that are highly overlapping, and inter-well transitions (fluxons) from ground state to 1\textsuperscript{st} excited state, exhibiting a vanishing overlap.
This spectrum classification is reminiscent of the fluxonium qubit~\cite{nguyen2019high}, which can also show a large anharmonicity.

As a summary, in contrast to the transmon, there is no need for a compromise between charge insensitivity and high anharmornicity in the interference-based $\cos(2\varphi)$ qubit when choosing the Josephson energy.
Both features become larger as $\ejsd/\ec$ increases, motivating the exploration of larger than usual $\ejsd/\ec$ ratio.
\label{text:larger}

\section{\texorpdfstring{$\bm{T_1}$}{T1} protection}

\subsection{How to estimate the decay?}

So far, we have only described the circuit protection against energy relaxation in a qualitative manner.
We will now move on to a quantitative discussion.
The circuit is described by two conjugate quantum operators, $\widehat{n}$ and $\widehat{\varphi}$ (see Fig.~\ref{fig:circuit}), for which the main associated relaxation mechanism are dielectric losses and coupling to the magnetic flux in the SQUID loop, respectively. The dominant source of decay  due to flux is expected to be 1/f flux noise picked up by the loop ~\cite{yan2016flux,bylander2011noise}.
To evaluate the relaxation rates and to be able to compare them to other quantum circuits, we first
calculate the relevant
matrix elements~\cite{Susskind1964, Koch2007}.

Dielectric loss couples via the charge operator
\begin{equation}
      \widehat{\mathcal{O}}_{\hat{n}} = 2e\hat{n}.
\end{equation}
The charge matrix element $\langle 0| \hat{n}  | 1 \rangle = M_n$ is thus the relevant figure of merit to discuss relaxation via the electric field.

For relaxation induced by magnetic flux noise, we derive the coupling operator
\begin{equation}
    \widehat{\mathcal{O}}_{\hat{\varphi}}  =  \frac{\partial \widehat{H} }{\partial \Phi}
\end{equation}
which using Eq.~\eqref{eq:ham2} gives
\begin{equation}
    \begin{aligned}
        \widehat{\mathcal{O}}_{\hat{\varphi}} = & \;\; - \ejsu \frac{\pi}{\Phi_0}
        \left[ \phantom{-}\cos\left(\hat{\varphi}\right) \cos\left(\pi \delta \Phi \right)
        \right.  \\
        & \;\;\;\;\;\;\;\;\;\;\;\;\;\;\;\;\; \left.+ d \sin\left(\hat{\varphi}\right) \sin\left(\pi \delta \Phi\right)  \right] \\
        & - \ejsd \frac{2\pi}{\Phi_0}
        \left[ -\cos\left(2\hat{\varphi}\right) \sin\left(2\pi \delta \Phi \right)\right.  \\
        & \;\;\;\;\;\;\;\;\;\;\;\;\;\;\;\;\; \left.+ d \sin\left(2\hat{\varphi}\right) \cos\left(2 \pi \delta \Phi \right) \right]
     \end{aligned}
\end{equation}
Several
operators are present involving either $1\hat{\varphi}$ or $2\hat{\varphi}$ harmonic terms. We consequently have two figures of merit:

\begin{itemize}[leftmargin=1pt]
    \item[] \resizebox{1\hsize}{!}{$M_{1 \varphi} = \phantom{-}\langle 0| \cos (\phantom{2} \widehat{\varphi})| 1 \rangle\cos(\phantom{2}\pi \delta \Phi) + \langle 0| \sin ( \phantom{2}\widehat{\varphi})| 1 \rangle d \sin(\phantom{2}\pi \delta \Phi) $
    }
    \item[] \resizebox{1\hsize}{!}{$M_{2 \varphi} = -\langle 0| \cos (2 \widehat{\varphi})| 1 \rangle \sin(2 \pi \delta \Phi) + \langle 0| \sin (2 \widehat{\varphi})| 1 \rangle d\cos(2 \pi \delta \Phi) $
    }
\end{itemize}
Notice their significantly different external magnetic field dependence
which may strongly influence the final relaxation rate.
Furthermore, the junction asymmetry, even as low as $d = \SI{1}{\percent}$ promotes interactions via the $\sin(n\,\widehat{\varphi})$ operators and will influence the final relaxation rate.
In the following sections, we will discuss quantitatively these figures of merit.

\subsection{Junction asymmetry}

\begin{figure}[ht]
    \centering
    \includegraphics[width=\linewidth]{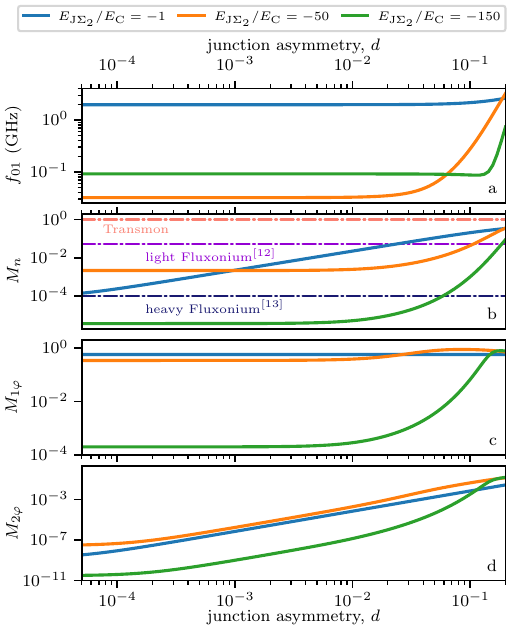}
    \caption{\textbf{Effect of the junctions asymmetry}.
    For $\ejsd/\ec \in \{-1, -50, -150\}$, we show the effect of the junction asymmetry $d$ on the qubit frequency, a), and its relevant relaxation figure of merits, b), c), d).
    In b), we show in pink and yellow dash-dot lines the typical matrix element for the transmon~\cite{Koch2007} and fluxonium~\cite{Somoroff2023, Earnest2018} qubit respectively.
    We see that for such a low asymmetry all figure of merits reach $<~10^{-2}$, confirming that the relaxation protection is well preserved for experimentally achievable asymmetries.
    For all figures, we have $\ejsd/\ejsu=-0.1$, $\delta \Phi = 10^{-5}$ and $n_\mathrm{g}=0.25$.
    }
    \label{fig:asymmetry}
\end{figure}

In this section, we examine how asymmetry in the junctions affects the matrix elements involved in qubit relaxation.
In Fig.~\ref{fig:asymmetry} we plot, for various $\ejsd/\ec$ ratio, the qubit frequency and relaxation figures of merit as function of the junction asymmetry $d$.
From panel b) to d), we see that to reach low matrix element values, i.e. long relaxation time, both low asymmetry and large $\ejsd/\ec$ are required.
Indeed, symmetry is required for the wavefunctions to localize in different wells and a large $\ejsd/\ec$ ratio increases the barrier height.
Quantitatively, the figure shows that the interference-based $\cos(2 \varphi)$ qubit can reach matrix element values several orders of magnitude lower than transmons~\cite{Koch2007} that can also be found in
fluxonium qubits~\cite{Somoroff2023,Earnest2018, Lin2018,ardati2024}, ensuring an efficient protection from energy relaxation.

\subsection{Flux \texorpdfstring{$\bm{\Phi}$}{phi} operating point}

\label{flux_operating_point}

\begin{figure}[ht]
    \centering
    \includegraphics[width=\linewidth]{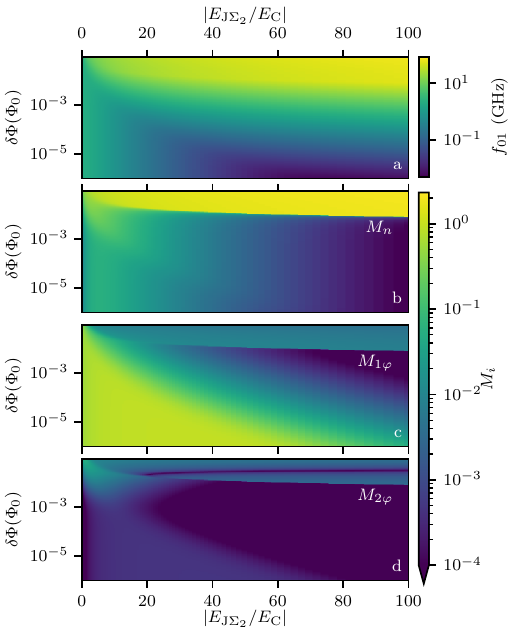}
    \caption{\textbf{Effect of magnetic flux offset.} a) Transition energy $f_{01}$ for varying magnetic flux and $\ejsd/\ec$ ratio. b) Charge matrix element $M_n$ for varying magnetic flux and $\ejsd/\ec$ ratio. c) $M_{1 \varphi}$ figure of merit flux matrix element for varying magnetic flux and $\ejsd/\ec$ ratio. d) $M_{2 \varphi}$ figure of merit for flux matrix element for varying magnetic flux and $\ejsd/\ec$ ratio.
    Bottom colorscale of b), c) and, d) were truncated to $1 \times 10^{-4}$ for the sake of visibility.
    For all figures, we have $\ejsd/\ejsu=-0.1$, $d = \SI{1}{\percent}$ and $n_\mathrm{g}=0.25$.}

    \label{fig:matrix-element}
\end{figure}

We now turn to the effect of the flux operating point on matrix element values.
Any realistic experiment will feature a small yet non-vanishing additional offset flux as well as flux noise so that $\delta \Phi \neq 0$, introducing again a $\cos(1\varphi)$ term in the Josephson potential.
We show in Fig.~\ref{fig:matrix-element} the resonance frequency and the figures of merit for energy relaxation in the different channels.
All plots separate into 3 regions.
In the left and bottom, $\ejsd/\ec$ is small resulting in significant inter-well coupling and hybridization of wavefunctions (see Fig.~\ref{fig:potential}.a).
In the bottom right, $\ejsd/\ec$ is large, dramatically reducing inter-well coupling and promoting in-well localization of the wavefunctions (see Fig.~\ref{fig:potential}.b - right).
In the top part of the figures, for large offset flux, the circuit leaves the protected regime as the first transition becomes intra-well, with a large wavefunction overlap (see Fig.~\ref{fig:potential}.b - left).

In Fig.~\ref{fig:matrix-element}.a), at low $\ejsd/\ec$, the resonance frequency is dominated by the inter-well coupling interaction and is mostly independent of $\delta \Phi$.
At high $\ejsd/\ec$, the resonance frequency is given by the imbalance between the two potential wells.
The discontinuity observed in panels b) to d) at $\delta \Phi \sim 10^{-2}$ comes from a level crossing between states $\ket{1}$ and $\ket{2}$.

Focusing on small offset flux, i.e. the bottom of the graphs, we observe in b) a striking property: the charge matrix element is almost independent of $\delta \Phi$ and is small in both the hybridized and the localized regimes.
This means that in an interference-based $\cos(2\varphi)$ qubits the energy relaxation due to charge matrix elements is resilient to significant offset flux $\delta \Phi$.
This is an unexpected feature. The vanishing energy relaxation is usually believed to be associated, in the $\cos(2\varphi)$ regime, to charge parity protection, i.e. that $\ket{0}$ and $\ket{1}$ states are built on disjoint supports (odd and even charge states).  This picture only holds for small offset flux, in the inter-well hybridized regime, i.e. the bottom left of the graph (see also Fig.~\ref{fig:potential}.a) bottom panel). This figure shows that the protection mechanism extends beyond this disjoint wavefunction picture: at larger offset flux, $\ket{0}$ and $\ket{1}$ both have contributions on even and odd charge states. Nevertheless, they are still localized in different wells, and their symmetry in charge space is such that the charge matrix element still vanishes. In Appendix B, we provide details about wavefunction symmetries in charge representation.  This explains the robustness of the protection mechanism.

In Fig.~\ref{fig:matrix-element}.c) and d), the figures of merit of flux-induced relaxation show overall an enhanced protection at large $\ejsd/\ec$ and increasing $\delta \Phi$.
Looking into details of the different figures of merit, the dominant operator for relaxation is $\cos(\hat{\varphi})$ whose symmetry enables a large overlap between inter-well hybridized wavefunctions.
This is unlike a typical transmon, in which the dominant term is $\sin(\hat{\varphi})$ due to intra-well transitions.
Shifting away from $\delta \Phi = 0$ reduces the phase matrix elements by promoting in-well localization of wavefunctions i.e. transitioning from an hybridized regime to a localized one.
``$2\varphi$'' matrix elements follow similar variations to ``$1\varphi$'' matrix elements, with however a decrease amplitude due to favorable weighting, i.e. vanishing $\sin(2\pi \delta \Phi)$ and small asymmetry factor $d$. In a general perspective, the in-well wavefunction localization enables the suppression of the matrix element associated to all decay mechanisms that couple through the phase degree of freedom.

To conclude, interference-based $\cos(2 \varphi)$ circuits show
vanishing charge and phase matrix element amplitudes even with finite offset flux $\delta \Phi$ and junction asymmetry $d$.
Larger $\ejsd/\ec$ leads to lower matrix element amplitudes, motivating the exploration of larger than usual $\ejsd/\ec$ ratio, as already seen in Sec.~\ref{text:larger}. Matrix elements smaller than $\sim 10^{-3}$ are achievable for both flux and charge, which is orders of magnitude better than transmons.

\section{\texorpdfstring{$\bm{T_{\varphi}}$}{Tphi} protection}

\subsection{How to estimate the dephasing?}

From the previous section, it is clear that our circuit holds protection against energy relaxation even when junction asymmetry and finite magnetic flux offset are taken into account.
We will now quantitatively discuss the dephasing of the qubit transition due to charge and flux noise.
To evaluate such dephasing and to be able to compare them to other quantum circuits, we compute~\cite{Ithier2005} $\braket{1| \partial_\lambda H|1} - \braket{0| \partial_\lambda H |0} \sim \partial_\lambda f_{01}$ for $\lambda \in \{ n_\mathrm{g}, \Phi \}$.

\begin{figure}[ht]
    \centering
    \includegraphics[width=\linewidth]{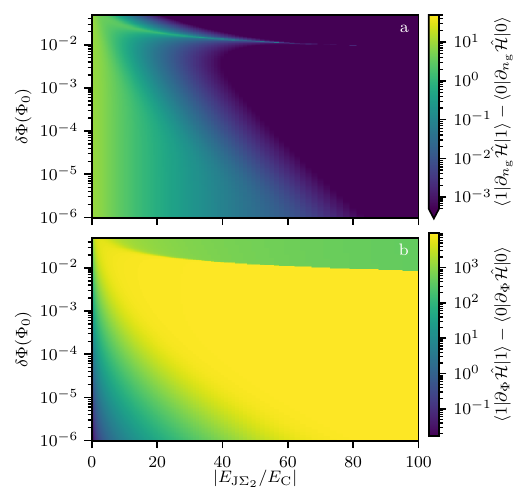}
    \caption{\textbf{Susceptibility to dephasing due to charge and flux noise. }
    Sensitivity of eigenstates energies to a) charge noise and b) flux noise. Lower bound of the color-scale of a) was truncated to $5\times10^{-4}$ for the sake of visibility.
    For all figures, we have $\ejsd/\ejsu=-0.1$, $d=\SI{1}{\percent}$ and  $n_\mathrm{g}=0.25$.
    \label{fig:dephasing}
    }
\end{figure}

\subsection{Trade-off between charge noise and flux noise}

A direct observation in Fig.~\ref{fig:dephasing} is the incompatibility between regions of insensitivity to charge noise, i.e. with large $\ejsd/\ec$ ratio, and to flux noise, requiring small $\ejsd/\ec$ ratio. This is one of the main conclusions of this work:
it is not possible to protect the interference-based $\cos(2 \varphi)$ qubits from both sources of dephasing.

In more detail, low $\ejsd/\ec$ ratio leads to significant charge dispersion, as in a transmon. The effect is however significantly enhanced in $\cos(2\varphi)$ qubits as the exponential coefficient of the charge dispersion suppression is twice lower~\cite{Smith2020}, which is a direct consequence of preventing single Cooper pair tunneling.
On the other hand, in the transmon regime at $\ejsd \gg \ec$, the suppression of the first order Josephson energy $\ejsu$ is controlled by interference between the SQUID's arms and is very sensitive to flux noise.
An $|\ejsd / \ejsu|$ ratio as large as possible reduces this flux noise sensitivity and is hence desirable. This matter is discussed quantitatively in section \ref{sec:ComprehensiveStudy}.

\subsection{\texorpdfstring{$\bm{T_{\varphi}}$}{Tphi}: inter-well coupling interpretation}

\begin{figure}[ht]
    \centering
    \includegraphics[width=\linewidth]{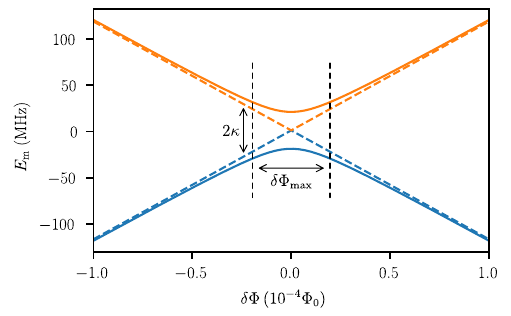}
    \caption{\textbf{Flux dispersion}.
    Ground state and first excited state energies versus offset flux for $n_g = 0$ (solid line) and $n_g = 0.5$ (dashed line).
    For $|\delta \Phi|  > \delta \Phi_{max}$, the flux dispersion is linear due to in-well localization of wavefunctions. For $|\delta \Phi|  < \delta \Phi_{max}$ in the center part, the two wells' wavefunctions hybridize for $n_g=0$.
    $\ejsd/\ejsu=-0.1$, $\ejsd/\ec = 40$ and $d=0$.
    }
    \label{fig:Sweetspot}
\end{figure}

We can understand these features by examining the consequences of inter-well coupling on the spectrum.
Fig.~\ref{fig:Sweetspot} shows the ground state and first excited state energies as function of offset flux $\delta \Phi$ for $n_g=0$ and $n_g=0.5$, solid and dashed line respectively.
Close to a pure $\cos(2\varphi)$ regime, $\delta \Phi \sim 0$, at $n_g =0$, the parabolic dispersion results from the inter-well interaction with interaction strength given by~\cite{Smith2020}:
\begin{equation}
    \kappa(n_g) = 8 \ec \sqrt{\frac{2}{\pi}} \left(\frac{2\ejsd}{\ec}\right)^{\frac{3}{4}} e^{-\sqrt{\frac{2\ejsd}{\ec}}} \cos(\pi n_g)
    \label{eq:kappa}
\end{equation} %
The inter-well interaction is
modulated by the charge offset $n_g$. When the interaction is finite
the
flux-dispersion
vanishes
at $\delta \Phi \sim0$
corresponding to
a sweet spot in flux.
However, we observe a very different behavior at $n_g = 0.5$, which is a high symmetry point for charge. Here the interaction vanishes, resulting in the absence of a flux sweet spot.  Consequently, the charge dispersion is large, as large as the transition frequency $f_{01}(n_g = 0)$ itself.

The robustness of the sweet spot regarding flux fluctuations relates to the second derivative versus $\delta \Phi$ of the qubit fundamental frequency. In a simplified
picture, the qubit frequency reads $f_{01} = 2 \sqrt{\kappa^2 + (\alpha \delta \Phi)^2 }$, where $\alpha = \pi |\ejsu| $ is the slope of the flux-induced degeneracy lifting between the two potential wells (without coupling). We define here the sweetness $\delta \Phi_\mathrm{max}$ which quantifies the sweet spot robustness:

\begin{equation}
    \delta \Phi_\mathrm{max} = \frac{\alpha}{\frac{\partial^2 f_{01}}{\partial \delta \Phi^2} \scriptstyle(\delta \Phi = 0)} = \frac{\kappa}{ \pi |\ejsu|}
    \label{eq:dphimax}
\end{equation}
$\delta \Phi_\mathrm{max}$ identifies to the range of offset flux $\delta \Phi$ for which flux dispersion is significantly suppressed. The value of $\delta \Phi_\mathrm{max}$ sets the boundary between two different regimes visible in Fig.~\ref{fig:Sweetspot}.
For $\delta \Phi < \delta \Phi_\mathrm{max}$ the two wells' wavefunctions hybridize due to inter-well coupling, and for $\delta \Phi > \delta \Phi_\mathrm{max}$ the wells degeneracy lifting is large enough to localize both wavefunctions.

Efficient transmon-like suppression of charge dispersion can be achieved by shifting away from the sweet spot with $\delta \Phi > \delta \Phi_{max}$ (see Fig.~\ref{fig:transmon-vs-cos2phi}), which remain compatible with the $T_1$ protection, but here the linear flux dispersion would make the qubit highly sensitive to flux noise, resulting in a small $T_{\varphi}$. Next we explore the possibility to find a suitable trade-off between the two regimes.

\section{Optimizing circuit parameters}
\subsection{Realistic experimental parameters}
\label{sec:real-params}

In this section, we explore the $\ejsd$ - $\ec$ - $\delta \Phi$ parameter space. We limit the study to experimentally relevant parameters for interference-based circuits.

The smallest acceptable value of $\ejsd$ is
obtained by comparing the thermal energy $k_B T$ with the depth
of the Josephson potential
$2\times \ejsd$. 
In practice, a circuit
with a photonic temperature $T \approx  \SI{50}{\milli\kelvin} \approx  \SI{1}{\giga\hertz}$~\cite{Somoroff2023} greater than $2\times \ejsd/k_B$ will be thermally excited to states of the continuum above the potential, and not usable as a qubit.
This effect is absent
in some other low-frequency qubits, such as the fluxonium qubits, with unbounded potential, for which confinement is always ensured by the inductive energy.

Very large values of $\ejsd$, up to several hundreds of GHz, are
accessible as critical current value of several µA have been demonstrated in graphene based Josephson junctions for instance~\cite{borzenets2016ballistic}.
On the other hand, achieving very large shunt capacitance is a
fabrication challenge,
which limits the lowest possible value of $E_C$ to several MHz~\cite{rousseau2025enhancing} using typically aluminum based planar superconducting circuits. Parallel plate capacitors could provide a way to overcome this limitation but remain limited due to the difficulty to deposit reproducibly thin films of lossless dielectric materials~\cite{wang2022hexagonal}. On the large $E_C$ side, capacitances can be reduced to few fF  in hybrid semiconductor Josephson junctions, enabling $\ec$ in the tens of GHz range.

Finally, realistically small and well defined offset flux $\delta \Phi$ is constrained by the minimum magnetic flux noise value $\sim ~ 10^{-6}\, \Phi_0$ in state of the art experiments~\cite{Ithier2005}, so that we choose $\delta \Phi = 10^{-5}$ as a minimum value in the following.

\subsection{Moderate \texorpdfstring{$\bm{\ejsd / \ec}$}{EJS2/EC} for minimum dephasing}
\label{sec:EjEc-compromise}

\begin{figure}[ht]
    \centering
     \includegraphics[width=\linewidth]{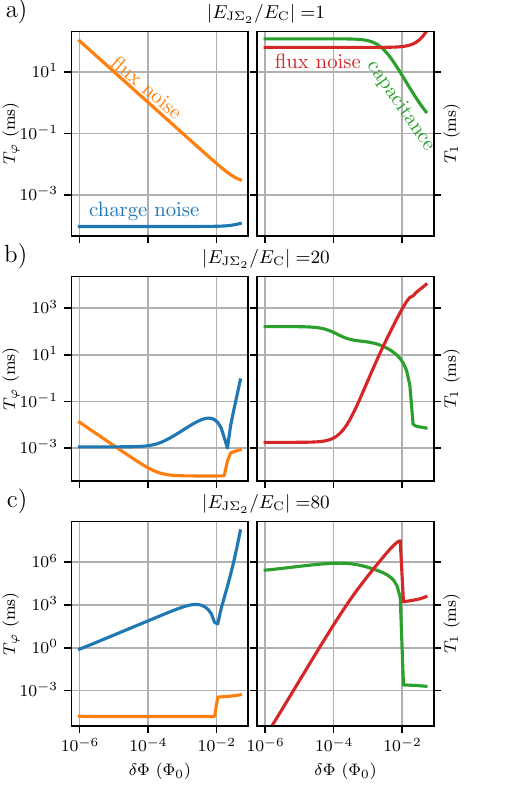}
    \caption{\textbf{Dephasing $\bm{T_{\varphi}}$ and decay $\bm{T_1}$ times dependence on $\bm{\delta \Phi}$ for different $\bm{\ejsd / \ec}$ ratio.} Junctions' asymmetry is fixed to $d=\SI{1}{\percent}$, the charging energy is set to $\ec =500$~MHz  and the ratio $\ejsd / \ec$ varies. It is equal to $ 1, 20, 80$, from top to bottom. The discontinuity observed for $\delta \Phi \geq 10^{-2}$ indicates the transition between the protected and unprotected regimes.}
    \label{fig:Tphi_vs_flux}
\end{figure}

Figure~\ref{fig:Tphi_vs_flux} shows estimations of pure dephasing times $T_{\varphi}$ (left) and decay times $T_1$ (right) as a function of flux offset $\delta \Phi$, for different $\ejsd / \ec$ ratio.
$T_{\varphi}$ estimations were obtained using typical $1/f$ noises as implemented in scQubits~\cite{Chitta_2022} with noise amplitudes $10^{-6}\, \Phi_0$ for flux noise and $10^{-4}\, e$ for charge noise~\cite{Smith2020}.
$T_1$ due to dielectric losses is estimated using a capacitance quality factor $Q_\mathrm{capa} = 10^{6}$~\cite{nguyen2019high}, and $T_1$ due to magnetic field relaxation is estimated using a noise amplitude of $10^{-6} \Phi_0$~\cite{nguyen2019high}.

As a general trend, we observe that the dephasing time $T_{\varphi}$ is more limiting than the decay time $T_{1}$, with $T_{\varphi}$ values never exceeding a few \SI{}{\micro\second}.
Focusing first on dephasing, we observe in panel a) at low $\ejsd/\ec$ ratio that charge noise dominates the decoherence as in a Cooper-pair-box-like regime, and $T_{\varphi}$ cannot exceed a few tens of nanoseconds.
At large $\ejsd/\ec$ ratio, in panel c), the linear dispersion versus flux dominates and $T_{\varphi}$ is limited to only a few nanoseconds due to flux noise.
In the intermediate regime of $\ejsd/\ec \sim 20 $, panel b), charge and flux noises both have effects of similar magnitude.
The effect of flux noise is
reduced by lowering $\delta \Phi$
until
charge noise
becomes the limiting factor.
We see that $T_{\varphi}$ has an upper limit on the order of \SI{1}{\micro\second}.
Hence, the most favorable configuration is to balance the effect of charge and flux noise, which happens for moderate $\ejsd/\ec$, see panel b).

Considering now $T_1$, we observe that $1/f$ flux noise is the dominant decay mechanism in the protected regime ($\delta \Phi \ll 10^{-2}$), which is expected since the charge
matrix element is vanishing by symmetry.
The plateaus of the $T_1$ versus flux is reminiscent of the robustness of the $T_1$ protection away from the flux sweet spot discussed in Sec.~\ref{flux_operating_point}.


\subsection{
Zero-frequency limits}
\label{sec:zero-freq-limit}

Here, we discuss the zero frequency limit regimes of operation and their practical applicability. Low frequency qubits are generally appealing to mitigate dephasing issues because dephasing rates often scale with the fundamental qubit frequency \cite{zhang2021universal}. There are two ways to obtain vanishingly low fundamental frequencies in $\cos(2\varphi)$ qubit.

\begin{figure}[ht!]
    \centering
     \includegraphics[width=\linewidth]{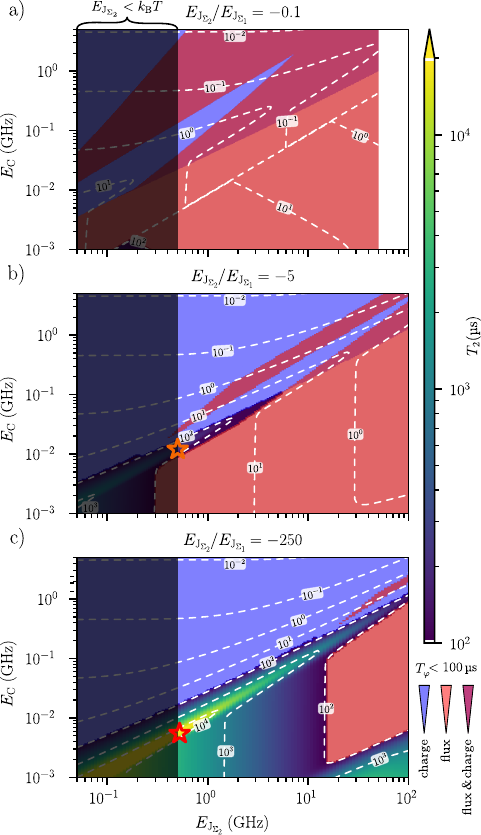}
    \caption{\textbf{Maximal coherence time $T_2$ achievable for given $\ec$ and $\ejsd$.}
    The junction asymmetry is fixed at $d = \SI{1}{\percent}$, and the charge offset is set to $n_\mathrm{g} = 0.25$. The ratio $\ejsd/\ejsu$ takes the values (a) $-0.1$, (b) $-5$, and (c) $-250$. For each point, the flux offset $\delta \Phi$ is varied in the range $10^{-5}$ to $9 \times 10^{-3}$ to maximize $T_2$ (the corresponding optimal $\delta \Phi$ values are provided in Appendix D).
    Colored regions indicate regimes of reduced qubit coherence due to different limiting mechanisms: temperature (black, where $2 \times \ejsd < k_B T / h \sim \SI{1}{\giga\hertz}$), charge noise (blue, where $T_\varphi^\mathrm{charge} < \SI{100}{\micro\second}$), and flux noise (salmon, where $T_\varphi^\mathrm{flux} < \SI{100}{\micro\second}$). For clarity, the color scale is truncated between $T_2 = \SI{100}{\micro\second}$ and $\SI{20}{\milli\second}$.
    Stars indicate the operating points that yield the longest coherence times, namely $(\ec, \ejsd)$ pairs corresponding to $T_2 = \SI{270}{\micro\second}$ for $\ejsd/\ejsu = -5$ and $T_2 = \SI{24}{\milli\second}$ for $\ejsd/\ejsu = -250$.
    }
    \label{fig:EJ2EJ1_limits}
\end{figure}

First, we discuss the regime $\delta \Phi = 0$ ($\Phi/\Phi_0 = 0.5$) and large $\ejsd / \ec\gg 50$. At $\delta \Phi = 0$, the fundamental qubit frequency is given by $f_{01} = 2 \kappa$ and the charge dispersion is also equal to $f_{01}$ (see Fig.~\ref{fig:Sweetspot}).
Similarly to the transmon,
increasing $\ejsd / \ec$
can be proposed to flatten the charge dispersion by reducing $\kappa$ exponentially (see Eq.~\eqref{eq:kappa}),
resulting in charge-insensitive qubit with
vanishingly small fundamental frequency.
Additionally, the sweet spot at $\delta \Phi = 0$ can be expected to ensure flux noise insensitivity. However, the sweetness $\delta \Phi_\mathrm{max}$ is reduced exponentially
with $\ejsd / \ec$ (see Eq.~\eqref{eq:dphimax} and \eqref{eq:kappa}), making the circuit highly sensitive to flux noise at the second order.
In practice, with a realistic $\ejsd / \ec = 70$, the charge dispersion remains as large as \SI{4}{\mega\hertz} and $\delta \Phi_\mathrm{max} \lesssim 10^{-6}\, \Phi_0$, i.e. smaller than the experimentally achievable flux noise amplitude. The magnetic flux sweet spot is then in practice non-existent in the large $\ejsd / \ec$ regime. This situation is illustrated in Fig.~\ref{fig:Tphi_vs_flux}.c): we do not observe any increase in $T_{\varphi}$ due to flux noise when $\delta \Phi$ decreases because the sweet spot is never reached, even at $\delta \Phi = 10^{-6}$.
For these reasons, relying on a degenerate computational space ($f_{01} \sim 0$) in the large $\ejsd/\ec$ limit is not realistic using the interference-based $\cos(2\varphi)$ qubit architecture.
More complex architecture based on an array of SQUIDs have been proposed to address this issue, providing enhanced extension of the sweet spot, see Ref.~\cite{schrade2022protected}.

A second regime of low frequency $\cos(2\varphi)$ qubits can be reached by reducing both $\ejsd$ and $\ec$ at constant, moderate values of the $\ejsd / \ec$ ratio, hence conserving an optimum regime regarding $T_{\varphi}$. This strategy is however limited as well, because of temperature since $\ejsd$ cannot be smaller than $k_B T /2 h \sim \SI{0.5}{\giga\hertz}$~\cite{Somoroff2023}.
We conclude that a very low frequency qubit, which was proposed as a mean to suppress dephasing \cite{zhang2021universal}, cannot increase the coherence time of $\cos(2\varphi)$ qubits to arbitrarily large values.

\subsection{Comprehensive study of circuit parameters}
\label{sec:ComprehensiveStudy}

\begin{table}[ht]
\centering
\begin{tabular}{lllllllll}
\toprule
$f_{01}$         & $\alpha$         & $\delta \Phi$ & $\ejsd$ & $\ec$   & $\ejsd$  & $T_1$           & $T_\varphi$     & $T_2$  \\
(MHz) & (MHz) & ($\Phi_0$)    & (GHz)         & (GHz) & $\overline{\ec\phantom{1\,\,}}$ & (\si{\micro\second}) & (\si{\micro\second}) & (\si{\micro\second}) \\
\midrule
$5.10^3$ & $53.10^{3}$ & $10^{-5}$&  48.5 & 5.6  & $\phantom{0}$9 &  $1.10^6$ & $\phantom{0}$0.02 & $\phantom{0}$0.02  \\
$1.10^3$ & $16.10^{3}$ & $10^{-5}$&  15.7 & 1.5  & 11 &  100 & $\phantom{0}$0.12 & $\phantom{0}$0.12  \\
$\phantom{0}$500 & $\,\,\,9.10^{3}$ & $10^{-5}$ & $\phantom{0}$8.6 & 0.8  & 11 &  $\phantom{0}$46 & $\phantom{0}$0.24 & $\phantom{0}$0.24  \\
$\phantom{0}$100 & $\,\,\,1.10^{3}$ & $10^{-5}$ & $\phantom{0}$1.2 & 0.1  & $\phantom{0}$9 &  $\phantom{0}$69 & $\phantom{0}$1 & $\phantom{0}$1  \\
\bottomrule
\end{tabular}
\caption{\textbf{Optimal $\bm{\ejsd}$ and $\bm{\ec}$ 
to obtain the longest coherence time $\bm{T_2}$. }
For all simulations, we used $\ejsd/\ejsu=-0.1$ and $n_\mathrm{g}=0.25$.
${\ejsd/\ec \sim 10}$ is consistently found to be the optimal ratio to maximize $T_2$.
Similarly the lowest allowed flux offset, $\delta \Phi = 10^{-5} \Phi_0$, always leads to the longest $T_2$.\newline
}
\label{tab:optimized}
\end{table}

We perform in this section a comprehensive numerical investigation of the circuit coherence as function of $\ejsd$, $\ec$ and $\delta \Phi$. 
We present in Appendix~\ref{app:C} the analytical method used to calculate coherence times in the low-frequency limit, that accounts for specific contributions of higher excited states due to thermal excitation. Detailed contributions of the different decoherence mechanisms are presented in Appendix~\ref{app:D}.

To start the discussion, we focus on the case
$\ejsd/\ejsu=-0.1$, which corresponds to realistic circuit parameters. We summarize in Table~\ref{tab:optimized} the longest
coherence times $T_2$ 
for different qubit frequencies $f_{01}$.
For a high qubit frequency
$f_{01}$ = \SI{5}{\giga\hertz}, the longest coherence is achieved for a moderate ratio $\ejsd / \ec \sim 9$, the limiting factor being the dephasing time $T_{\varphi}$ with similar contributions of charge and flux noise.
As $f_{01}$ is reduced, the optimal $\ejsd / \ec$ ratio shows little variations. The dephasing time $T_{\varphi}$ increases as 1/f, while $T_1$ decreases but
remains
much larger, leading to an overall longer $T_2$. 
We conclude that
interference-based $\cos(2\varphi)$ qubit
with
$\ejsd/\ejsu=-0.1$
leads, at best, to a coherence time $T_2$ of the order of a few microseconds.
This qubit is intrinsically limited by charge noise at low $\ejsd/E_{C}$ and flux noise at large $\ejsd/E_{C}$,
with
temperature setting a lower boundary for $\ejsd$.

To summarize this, in Fig.~\ref{fig:EJ2EJ1_limits}, we show the expected $T_2$
as a function of circuit parameters and for the optimal flux offset $\delta \Phi$
at each $\ejsd$, $E_C$.
Additionally, we represent by color shaded areas the main limiting factor for $T_2$:  charge noise (blue), flux noise (salmon) or temperature (black).
For $\ejsd/\ejsu = -0.1$ (see panel a)), the overlap between the different shaded regions illustrates that there is no satisfactory working point because of these three decoherence channels.

Finding acceptable circuit parameters requires extending the discussion to $|\ejsd/\ejsu|$ ratios larger than 1.
In panel b), we present the results for a ratio of 5, that could be obtained in the Flowermon circuit for instance (see Table \ref{tab:all-circuits}). In panel c), we explore the situation for a ratio as high as 250, which is currently unreachable in existing architectures.
We observe that increasing the $|\ejsd/\ejsu|$ ratio is favorable in terms of flux noise induced decoherence because $\partial_{\phi} f_{01} \propto \ejsu$ in the in-well localized regime.
This opens a domain with coherence times higher than \SI{100}{\micro \second}, culminating to $T_2=\SI{270}{\micro\second}$ [orange star of panel b)] and $T_2=\SI{24}{\milli \second}$ [red star of panel c)]. This is only a
small improvement compared to
state-of-the-art transmons \cite{Bland2025,Tuokkola2024,Dane2025} considering that it would require to increase $|\ejsd/\ejsu|$ by two orders of magnitude compared to what is currently achievable.



\section{Conclusion}

The concept of the $\cos(2\varphi)$ qubit offers an appealing pathway toward mitigating the energy relaxation mechanisms that limit all superconducting qubits. However, our analysis reveals that in interference-based implementations, this protection entails an unavoidable trade-off between flux and charge noise, which ultimately results in pure dephasing of the qubit. Interestingly, we have shown that these competing dephasing channels can be partially balanced by operating the qubit slightly away from its flux sweet spot.

By systematically exploring the parameter space of a generic circuit encompassing a broad range of experimental realizations, we find that with currently achievable circuit parameters, the dephasing time is limited to a few microseconds—substantially below the state-of-the-art coherence times demonstrated in transmons.

Looking ahead, we estimate that even for ratios as large as $\ejsd/\ejsu \sim -250$, corresponding to a two-order-of-magnitude improvement compared to any existing or proposed circuit, coherence times could reach approximately $20$ ms. This projection underscores the fundamental constraints of interference-based $\cos(2\varphi)$ architectures and raises important questions about their suitability as building blocks for scalable superconducting quantum processors.

These findings highlight the need for alternative design strategies to achieve simultaneous protection against energy relaxation and dephasing. One promising direction is the recently proposed interference-based architecture featuring a finite junction asymmetry with $\ejsd < 0$~\cite{Hays2025}. Such a configuration could, in principle, circumvent the trade-off identified in our work. Nonetheless, the experimental demonstration of a circuit element with $\ejsd < 0$ that remains compatible with high-coherence operation is an open and critical challenge for future research.

\section{Acknowledgment}

We acknowledge discussions with Manuel Houzet, Julia Meyer, Zaki Leghtas, and Erwan Roverc'h. This work was supported by the French National Research Agency (ANR) in the framework of the Graphmon project (ANR-19-CE47-0007) and the Grenoble LaBEX LANEF. This work benefited from a French government grant managed by the ANR agency under the ``France 2030 plan'', with reference ANR-22-PETQ-0003. J.-S. T. acknowledges support from Direction Générale de l’Armement.

\section{Data Availability}

The data that support the findings of this article are openly available \cite{messelot_2025_17909031}.

\appendix

\renewcommand\thefigure{S\arabic{figure}}
\setcounter{figure}{0}

\section{Cooper-pair pairing by interferences}
\label{app:A}

The circuit discussed in the main text is based on the interference between two branches that form a loop, which is threaded by approximately half a magnetic flux quantum. In this appendix, we explore different experimental implementations of the $\cos(2 \varphi)$ qubit. We demonstrate that they share the same Hamiltonian but with different energy scales.

\subsection{Interferences}

The inductive potential associated with a single branch can be expanded as a Fourier series
\begin{align*}
    \widehat{U}(\phiop) &= \sum_{m>1} E_{\mathrm{J}_m} \cos (m\phiop) \\
    &= \eju \cos (\phiop) + \ejd\cos (2\phiop) + ...
\end{align*}

Considering the two interfering branches, A and B, we get the effective total potential that pairs even number of Cooper pairs:
\begin{align*}
    \widehat{\widetilde{U}}(\phiop_A) &= \widehat{U}_A(\phiop_A) + \widehat{U}_B(\phiop_A + \pi) \\ & = \sum_{m\in 2 \mathbb{N}^*} E_{\mathrm{J}_m} \cos (m\,\phiop_A) \\
    &= \ejsd \cos (2\phiop_A) + ...
\end{align*}

Introducing an asymmetry $d_1$ such that the first harmonic energies of each branch are $E_{\mathrm{J}_A}=\frac{1-d_1}{2}\ejsu$ and $E_{\mathrm{J}_B}=\frac{1+d_1}{2}\ejsu$, the approximate effective potential is modified according to:
\begin{equation*}
    \widehat{\widetilde{U}}(\phiop_A) = -\ejsu d_1 \cos(\phiop_A) + \ejsd \cos (2\phiop_A) + ...
\end{equation*}

This equation corresponds to Eq.~\eqref{eq:ham2} in the case $\delta\Phi = 0$ with the change of variable $\phiop = \phiop_A + \frac{\pi}{2}$. 
Using these two formalisms, we can get a common description of all interference-based $\cos(2\varphi)$ qubits.

In the remainder of this Appendix, we give the expressions and orders of magnitude of $\ejsu$ and $\ejsd$ for different implementations of the $\cos(2 \varphi)$ qubit. The formalism used for a given equation will be expressed by the choice of variable, either $\phiop_A$ or $\phiop$.

\subsection{Rhombus}

The Rhombus is a device in which two branches of current-biased Josephson junctions interfere~\cite{Gladchenko2008,Bell2014}. On one branch, we have two junctions (a and b) in series:
\begin{align*}
  \widehat{U}(\phiop_A) &= -E_{\mathrm{J}_a} \cos (\phiop_a) - E_{\mathrm{J}_b} \cos (\phiop_A-\phiop_a) \\
  &= -E_{\mathrm{J}_a} \cos (\phiop_a) \\ &\quad\, - E_{\mathrm{J}_b} [\cos(\phiop_A)\cos(\phiop_a)+\sin(\phiop_A)\sin(\phiop_a)]
\end{align*}

Current conservation laws give:
$$E_{\mathrm{J}_a} \sin (\phiop_a) = E_{\mathrm{J}_b} \sin (\phiop_A-\phiop_a)$$


Introducing $\displaystyle \uptau = \frac{4E_{\mathrm{J}_a}E_{\mathrm{J}_b}}{(E_{\mathrm{J}_a}+E_{\mathrm{J}_b})^2}$, $\displaystyle E_{\mathrm{J}} = \frac{E_{\mathrm{J}_a}+E_{\mathrm{J}_b}}{2}$, and using trigonometric relationships, it simplifies to~\cite{alvise}:
$$\widehat{U}(\phiop_A) = -2E_{\mathrm{J}} \sqrt{1-\uptau \sin^2 \frac{\phiop_A}{2}}$$

Moreover, introducing the asymmetry $\eta$ such that $E_{\mathrm{J}_a} = (1+\eta)E_{\mathrm{J}}$, $E_{\mathrm{J}_b} = (1-\eta)E_{\mathrm{J}}$ gives
$$\uptau = 1-\eta^2$$
and
$$\widehat{U}(\phiop_A) = -2E_{\mathrm{J}} \sqrt{\cos^2 \frac{\phiop_A}{2}+\eta^2\sin^2 \frac{\phiop_A}{2}}$$

Assuming a small asymmetry $\eta \simeq 0$ i.e. $\uptau \simeq 1$, it is possible to write:
\begin{align*}
    \widehat{U}(\phiop_A) =& -2E_{\mathrm{J}} \left| \cos\frac{\phiop_A}{2} \right| \sqrt{1+\eta^2\tan^2\frac{\phiop_A}{2}} \\
    \simeq& -2E_{\mathrm{J}} \left| \cos\left(\frac{\phiop_A}{2}\right) \right| \left(1+\frac{\eta^2}{2}\tan^2\left(\frac{\phiop_A}{2}\right)\right) \\
    \simeq& -\frac{2 E_{\mathrm{J}} \left(2+\eta^2 (\log(4)-1)\right)}{\pi} \\
    &-\frac{4 E_{\mathrm{J}} \left(2 - \eta^2 (\log (64)-5)\right) \cos(\phiop_A)}{3\pi} \\
    &+\frac{2 E_{\mathrm{J}} \left(4 - 3 \eta^2 (5 \log (16)-26)\right) \cos (2\phiop_A)}{15\pi}
\end{align*}

The ratio $\ejsd / \ejsu$ is maximal for $\eta=0$ and equals $-1/5$.

\subsection{High transparency Josephson junctions}
Compared to tunnel junctions, high transparency Josephson junctions involved less transmitting channels with transparencies that can reach values close to 1. Such junctions have been implemented using graphene~\cite{Messelot2024a}, quantum-well hetero-structures~\cite{Leblanc2025}, pinhole Josephson junctions~\cite{Griesmar2025} or InAs nanowire~\cite{spanton2017current}.

In such junctions, in the short junction limit, for one dimensional channels, the potential can be written:
$$\widehat{U}(\phiop_A) = -N\Delta \sqrt{1 -\uptau\sin^2(\phiop_A/2)}$$
where $\uptau \in [0,1]$ is the channel transparency, N the number of channels and $\Delta$ the superconducting gap. It gives results similar to the Rhombus. The ratio $\ejsd / \ejsu$ depends on transparency and hence on the material. In practice it typically ranges from $-1/5$ to $-1/10$.

\subsection{KITE with a small inductance}

The Kinetic Interference coTunneling Element (KITE) is a circuit where each branch of the SQUID is made of a Josephson junction in series with an inductor (made as a chain of Josephson junctions). We can consider $E_{\mathrm{L}} \gg E_{\mathrm{J}}$~\cite{alvise}. A semi-classical approximation (Born-Oppenheimer approximation) gives:
\begin{equation*}
    \widehat{\widetilde{U}}(\phiop) = -2 E_{\mathrm{J}} \cos(\phiop)\cos\left(\frac{\phi_\textrm{ext}}{2}\right) - \frac{1}{2}\frac{E_{\mathrm{J}}^2}{E_{\mathrm{L}}}\cos(2\phiop)\sin^2\left(\frac{\phi_\textrm{ext}}{2}\right)
\end{equation*}

Thus $\displaystyle \frac{\ejsd}{\ejsu}=-\frac{E_{\mathrm{J}}}{4E_{\mathrm{L}}}$; since $E_{\mathrm{J}}$ is assumed to be much smaller than $E_{\mathrm{L}}$, we consider $E_{\mathrm{J}}/E_{\mathrm{L}}=0.1$ which gives a ratio $-1/40$.

\subsection{KITE with a large inductance}

We can also consider a KITE where $E_{\mathrm{L}} \ll E_{\mathrm{J}}$~\cite{Smith2020}.
Using scQubits~\cite{Chitta_2022}, we can compute the charge dispersion of this circuit even for a very large Josephson energy. Fitting this charge dispersion with the Hamiltonian~\eqref{eq:ham2}, we get the effective $\ejsu$ and $\ejsd$ of this system: the ratio is always between -1 and 0.

\subsection{Flowermon}
Using twisted cuprate hetero-structures~\cite{flowermon}, it is possible to create a Josephson junction with many harmonics. The amplitude of the first harmonic depends on the twist angle $\theta$ between the two flakes:
$$\widehat{U}(\phiop_A) = -E_{\mathrm{J}} \cos(2\theta) \cos \phiop_A + E_\kappa \cos(2\phiop_A)$$
with a typical ratio $E_\kappa/E_{\mathrm{J}}=0.1$.

\begin{figure}[ht]
    \centering
    \includegraphics[width=\linewidth]{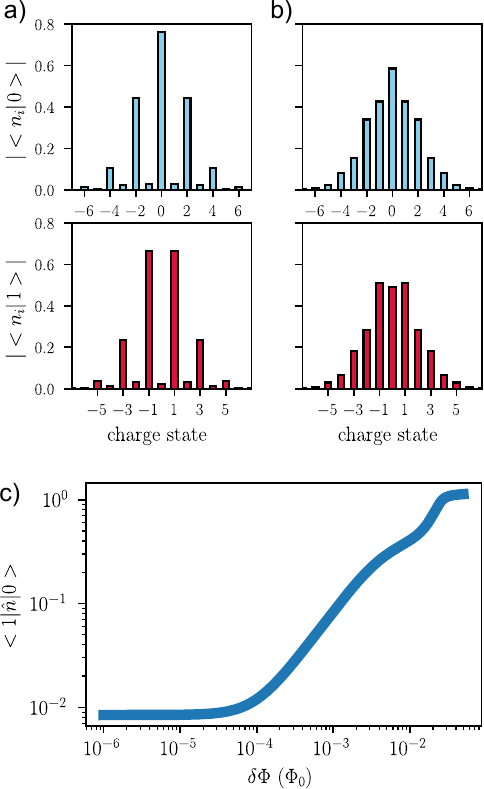}
    \caption{\textbf{States parity and charge matrix element}. \textbf{a)} Wavefunctions of the $\cos(2\varphi)$ qubit in charge representation for states $\ket{0}$ (blue) and $\ket{1}$ (red), at $\delta \Phi = 10^{-6}$.  \textbf{b)} Same at $\delta \Phi = 10^{-4}$. \textbf{c)} Charge matrix element of the computational states versus offset flux $\delta \Phi$. For all panels, $|\ejsd/E_C | = 40 $ }
    \label{fig:Figure10}
\end{figure}

Two branches of twisted cuprates can be associated in a SQUID~\cite{biflowermon}, with a simplified potential in the absence of asymmetry:

\begin{align*}
    \widehat{\widetilde{U}}(\phiop) \simeq&\ -2E_{\mathrm{J}}\cos(2\theta)\cos\left(\frac{\phi_\mathrm{ext}}{2}\right)\cos(\phiop) \\
    &+ 2E_\kappa\cos(\phi_\mathrm{ext})\cos(2\phiop)
\end{align*}

The ratio of the two harmonic depends on $\theta$ whose ideal working point is \SI{45}{\degree}, but experimentally we can expect at best the accuracy on the angle to be \SI{0.2}{\degree}:
$$\frac{\ejsd}{\ejsu}=\frac{E_\kappa}{E_{\mathrm{J}}\cos(2\theta)}$$
which gives a range of values $[-15, -0.1]\cup[0.1,15]$.

\section{Robustness of charge matrix element to offset flux \texorpdfstring{$\bm{\delta \Phi}$}{delta Phi} and wavefunction symmetry in charge space}
\label{app:B}
In this part, we provide an explanation for the relative insensitivity of the charge matrix element between $\ket{0}$ and $\ket{1}$ states of the $\cos(2\varphi)$ qubit to the offset flux $\delta \Phi$, that we observed in Fig.~\ref{fig:matrix-element}.b). For this purpose, we investigate the evolution of the wavefunctions in the charge basis representation and provide qualitative arguments based on symmetries and hybridization of wavefunctions.

\subsection{Robustness of charge matrix element suppression to offset flux \texorpdfstring{$\bm{\delta \Phi}$}{delta Phi}}

Fig.~\ref{fig:Figure10}.a) represents the absolute values of the projection of eigenstates $\ket{0}$ and $\ket{1}$ on the charge basis, at $\delta \Phi = 10^{-6}$. We observe that $\ket{0}$ has sizable components only on the even charge states, and $\ket{1}$ on the odd charge states. This way, $\ket{0}$ and $\ket{1}$ form the so-called ``parity-protected qubit" manifold, for which transition probabilities due to charge matrix element vanish due to a non-overlapping charge support.
Small additional contributions of the opposite parity arise from the finite offset flux and asymmetry (set to 1\%) between the 2 junctions.

The same representation at $\delta \Phi = 10^{-4}$ shows a very different picture in Fig.~\ref{fig:Figure10}.b): opposite parity contributions have increased with $\delta \Phi$ up to the point that we cannot observe any particular charge parity for $\ket{0}$ or $\ket{1}$ states. At $\delta \Phi = 10^{-4}$, the basic picture of parity protection does not hold.

We represent in Fig.~\ref{fig:Figure10}.c) the charge matrix element $\bra{1}\widehat{n}\ket{0}$ for the same circuit parameters, as a function of offset flux. From Fig.~\ref{fig:Figure10}.a) and b), we would expect a large increase of the charge matrix element but we see that the charge matrix element increases only moderately between $\delta \Phi = 10^{-6}$ and $\delta \Phi = 10^{-4}$, and the system remains in the protected regime. This observation shows that the simple picture presented before is not sufficient to explain the vanishing energy relaxation in this $\cos(2\varphi)$ qubit circuit.

\subsection{Symmetry of wavefunctions in charge basis}

\begin{figure}[ht]
    \centering
    \includegraphics[width=\linewidth]{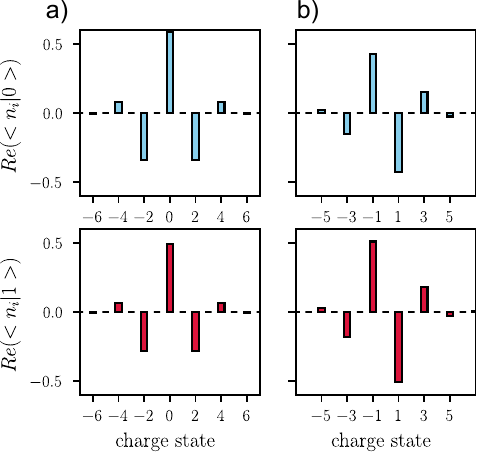}
    \caption{\textbf{ States symmetry and charge parity}. \textbf{a)} Even charge states part of the wavefunctions of states $\ket{0}$ (blue) and $\ket{1}$ (red), at $\delta \Phi = 10^{-4}$. \textbf{b)} Odd charge states part.}
    \label{fig:Figure11}
\end{figure}
To explain the persistence of protection at finite $\delta \Phi$, we investigate the symmetries of the eigenstates wavefunction in charge basis. In Fig.~\ref{fig:Figure11}, we show the real part of the projection of eigenstates $\ket{0}$ and $\ket{1}$ on the charge basis with the same circuit parameters as Fig.~\ref{fig:Figure10} and $\delta \Phi =10^{-4}$ (the imaginary part is one order of magnitude smaller). We separated even and odd charge states that we note $\ket{i}_{e}$ and $\ket{i}_{o}$ ($i=0,1$) in sub-figures a) and b) respectively to improve visibility of symmetries, but we have $\ket{i} = \ket{i}_{e} + \ket{i}_{o}$ as expected. Also, since the charge operator $\widehat{n}$ conserves parity, we can restrict the symmetry analysis to a given parity so that the charge matrix element reads:

\begin{equation}
    \bra{1}\widehat{n}\ket{0} = \bra{1}_{e}\widehat{n}\ket{0}_{e} + \bra{1}_{o}\widehat{n}\ket{0}_{o}
\end{equation}\

Fig.~\ref{fig:Figure11} shows that in $\cos(2\varphi)$ qubits, the even charge part of the wavefunction is symmetric versus charge inversion and the odd charge part is antisymmetric for both the ground and the first excited states. Using symbolic notation $A$ and $S$ for the symmetry of elements, we have ($\widehat{n}$ is antisymmetric):

\begin{equation}
    \bra{1}\widehat{n}\ket{0} = S|A|S + A|A|A
\end{equation}\

So that $\bra{1}\widehat{n}\ket{0} \approx 0$ from basic selection rules derived from symmetry. We show in this manner that the vanishing matrix element results from wavefunction symmetries in addition to a separation of wavefunctions by parity. This in turn explains the robustness of the charge matrix element suppression to finite offset flux $\delta \Phi$.\\

As a reminder in transmons, $\ket{1}$ and $\ket{0}$ have opposite symmetries in charge basis, hence the charge matrix element $\bra{1}\widehat{n}\ket{0}$ is large. This comes from the fact that transmons states are encoded in the symmetry inside a single potential well, whereas in $\cos(2\varphi)$ qubits, the $\ket{1}$ and $\ket{0}$ states are encoded in the two wells of the $\cos(2\varphi)$ potential. This highlights the essential importance of an additional degree of freedom in protected qubit schemes.

\section{Impact of higher excited states and temperature}
\label{app:C}
At finite temperature, the presence of higher excited states can create additional decoherence and decay channels. In this Appendix, we derive the expression for the rate associated with these processes.

We consider a system with computational states $|0\rangle$ and $|1\rangle$, together with $n-1$ higher levels. We describe this system using a density matrix that accounts for the possibility of coherence only between the first two levels.
\begin{equation*}
    \dm = \begin{pmatrix}
        p_0 & \rho_{01}&0 & \cdots & 0 \\
        \rho_{01}^* & p_1 & 0 & \cdots & 0 \\
        0 & 0 & p_2 & \cdots & 0 \\
        \vdots & \vdots & & \ddots & \vdots \\
        0 & 0 & 0 & \cdots & p_n
    \end{pmatrix}
\end{equation*}

The transition rate from state $i$ to state $j$ is $\rate{i}{j}$, which represents either excitation or de-excitation. This rate can be calculated using Fermi's golden rule. Associated to these rates, the jump operators can be written as  $L_{ij}=\sqrt{\rate{i}{j}} \ketbra{j}{i}$.

Thus, we compute the dissipation super-operators:
\begin{align*}
\mathcal{D}[L_{ij}](\dm) &= L_{ij} \dm L_{ij}\dag - \half\left(L_{ij}\dag L_{ij}\dm + \dm L_{ij}\dag L_{ij}\right) \\
&= \rate{i}{j} \left[p_i\ket{j}\!\!\bra{j} - \half \left(\ketbra{i}{i}\dm + \dm\ketbra{i}{i} \right) \right]
\end{align*}

From which we get the Lindblad master equation in the Heisenberg picture:
\begin{equation*}
    \dot{\dm} = \sum_{i,j\leq n} \rate{i}{j} \left[p_i\ketbra{j}{j}-\half \left( \ketbra{i}{i}\dm + \dm\ketbra{i}{i} \right) \right]
\end{equation*}



We thus extract the rate equations for the population:
\begin{equation}
    \dot{p}_k = \braket{k|\dot{\dm}|k} = \sum_{i\neq k} \rate{i}{k}\, p_i - \sum_{i\neq k} \rate{k}{i}\, p_k
    \label{eq:sde_populations}
\end{equation}

\subsection{Qubit decoherence rate}
We can extract the dynamics of the off-diagonal element of the density matrix:
\begin{align*}
    \dot{\rho}_{01} &= \braket{0|\dot{\dm}|1} = -\half \left[ \sum_{j\leq n} \rate{0}{j} + \sum_{j\leq n} \rate{1}{j}\right] \rho_{01} \\
    &\triangleq - \Gamma_2\, \rho_{01}
\end{align*}

We thus find the expression of the higher-state-induced decoherence:
\begin{equation*}
    \Gamma_2 = \frac{\Gamma_1^{01}}{2} +\frac{1}{2} \left[ \sum_{1<j\leq n} \rate{0}{j} + \sum_{1<j\leq n} \rate{1}{j}\right]
\end{equation*}
where $\Gamma_1^{01} = \rate{0}{1} + \rate{1}{0}$. Any additional dephasing mechanisms affecting the qubit, such as flux noise or photon shot noise, will contribute additively to this dephasing rate. In the simulations, we found that the contribution of temperature to dephasing was negligible, so we disregarded it.


\begin{widetext}
\subsection{Decay rate}
Equation~\eqref{eq:sde_populations} can be written in matrix form:
\begin{equation*}
    \begin{pmatrix}
        \dot{p}_0\vphantom{-\sum_{i\neq 0}} \\ \dot{p}_1\vphantom{-\sum_{i\neq 0}} \\ \vdots\vphantom{-\sum_{i\neq 0}} \\ \dot{p}_n\vphantom{-\sum_{i\neq 0}}
    \end{pmatrix}
    =
    \begin{pmatrix}
        -\sum_{i\neq 0} \rate{0}{i}      & \rate{1}{0}                 & \cdots & \rate{n}{0} \\
        \rate{0}{1}                      & -\sum_{i\neq 1} \rate{1}{i} & \cdots & \rate{n}{1} \\
        \vdots\vphantom{-\sum_{i\neq 0}} & \vdots                      & \ddots & \vdots \\
        \rate{0}{n}                      & \rate{1}{n}                 & \cdots & -\sum_{i\neq n} \rate{n}{i} \\
    \end{pmatrix}
    \begin{pmatrix}
        p_0\vphantom{-\sum_{i\neq 0}} \\ p_1\vphantom{-\sum_{i\neq 0}} \\ \vdots\vphantom{-\sum_{i\neq 0}} \\ p_n\vphantom{-\sum_{i\neq 0}}
    \end{pmatrix}
\end{equation*}

The excitation processes to the states out of the computational space being only due to thermal excitations, the excitation rates ($\rate{i}{j}$ with $i<j$) are much smaller than the decay rates ($\rate{j}{i}$ with $i<j$). Therefore, we consider a steady-state approximation, which translates to $\dot{p}_k = 0$ for $k \geq 2$. We obtain a block-matrix equation:

\begin{equation*}
    \begin{pmatrix}
        \dot{p}_0\vphantom{-\sum_{i\neq 0}} \\ \dot{p}_1\vphantom{-\sum_{i\neq 0}} \\ 0\vphantom{-\sum_{i\neq 0}} \\ \vdots\vphantom{-\sum_{i\neq 0}} \\ 0\vphantom{-\sum_{i\neq 0}}
    \end{pmatrix}
    =
    \underbrace{\left(\begin{array}{*2{>{\displaystyle}c}|*3{>{\displaystyle}c}}
        -\sum_{i\neq 0} \rate{0}{i}      & \rate{1}{0}                 & \rate{2}{0}                 & \cdots & \rate{n}{0}             \\
        \rate{0}{1}                      & -\sum_{i\neq 1} \rate{1}{i} & \rate{2}{1}                 & \cdots & \rate{n}{1}              \\
        \hline
        \rate{0}{2}                      & \rate{1}{2}                 & -\sum_{i\neq 2} \rate{2}{i} & \cdots & \rate{n}{2}                \\
        \vdots\vphantom{-\sum_{i\neq 0}} & \vdots                      & \vdots                      & \ddots & \vdots                      \\
        \rate{0}{n}                      & \rate{1}{n}                 & \rate{2}{n}                 & \cdots &  -\sum_{i\neq n} \rate{n}{i} \\
    \end{array}\right)}_{
    \left(\begin{array}{c|c}
         A & B \\
         \hline
         C & D
    \end{array}\right)
    }
    \begin{pmatrix}
        p_0\vphantom{-\sum_{i\neq 0}} \\ p_1\vphantom{-\sum_{i\neq 0}} \\ p_2\vphantom{-\sum_{i\neq 0}} \\ \vdots\vphantom{-\sum_{i\neq 0}} \\ p_n\vphantom{-\sum_{i\neq 0}}
    \end{pmatrix}
\end{equation*}
\end{widetext}

From this we get:
\begin{equation}
    \begin{pmatrix}
        \dot{p}_0 \\ \dot{p}_1
    \end{pmatrix}
    =
    A
    \begin{pmatrix}
        p_0 \\ p_1
    \end{pmatrix}
    +
    B
    \begin{pmatrix}
        p_2 \\ \vdots \\ p_n
    \end{pmatrix}
    \label{eq:sed_block}
\end{equation}
and
\begin{equation}
    \begin{pmatrix}
        0 \\ \vdots \\ 0
    \end{pmatrix}
    =
    C
    \begin{pmatrix}
        p_0 \\ p_1
    \end{pmatrix}
    +
    D
    \begin{pmatrix}
        p_2 \\ \vdots \\ p_n
    \end{pmatrix}
    \label{eq:inv_pop}
\end{equation}
Using equation~\eqref{eq:inv_pop}, the population of higher states can be expressed depending on $p_0$ and $p_1$:
\begin{equation*}
    \begin{pmatrix}
        p_2 \\ \vdots \\ p_n
    \end{pmatrix}
    = - D^{-1} C
    \begin{pmatrix}
        p_0 \\ p_1
    \end{pmatrix}
\end{equation*}

We can inject this result into the previous equation~\eqref{eq:sed_block} and obtain the effective transition rates in the computational subspace:
\begin{equation*}
    \begin{pmatrix}
        \dot{p}_0 \\ \dot{p}_1
    \end{pmatrix}
    =
    \underbrace{\left(A - B D^{-1} C\right)}_\Lambda
    \begin{pmatrix}
        p_0 \\ p_1
    \end{pmatrix}
\end{equation*}

This defines a $2 \times 2$ matrix
\begin{equation*}
    \Lambda = \begin{pmatrix}
        \Lambda_{00} & \Lambda_{01}\\
        \Lambda_{10} & \Lambda_{11}
    \end{pmatrix}
\end{equation*}

The decay rate of the qubit is thus given by
\begin{equation*}
    \Gamma_1 = \rate{0}{1}^\mathrm{eff}+\rate{1}{0}^\mathrm{eff}
\end{equation*}

with $\rate{0}{1}^\mathrm{eff} = \Lambda_{10}$ and $\rate{1}{0}^\mathrm{eff}=\Lambda_{01}$ the effective excitation and de-excitation rates of the qubit, respectively.



\subsubsection*{Application to the case n = 2}
We now give the explicit expression of the decay ($\Gamma_1$) and decoherence ($\Gamma_2$) rates in the presence of a single higher level. One finds the following:
\begin{align*}
    &\displaystyle\Gamma_1 = \rate{0}{1} + \rate{1}{0} + \frac{\rate{1}{2} \rate{2}{0} + \rate{0}{2} \rate{2}{1}}{\rate{2}{0} + \rate{2}{1}}\\
    &\displaystyle\Gamma_2 = \frac{\rate{0}{1} + \rate{1}{0} + \rate{0}{2} + \rate{1}{2}}{2}
\end{align*}

\section{Exhaustive circuit parameters study}
\label{app:D}

\begin{figure}[ht!]
    \centering
    \includegraphics[width=\linewidth]{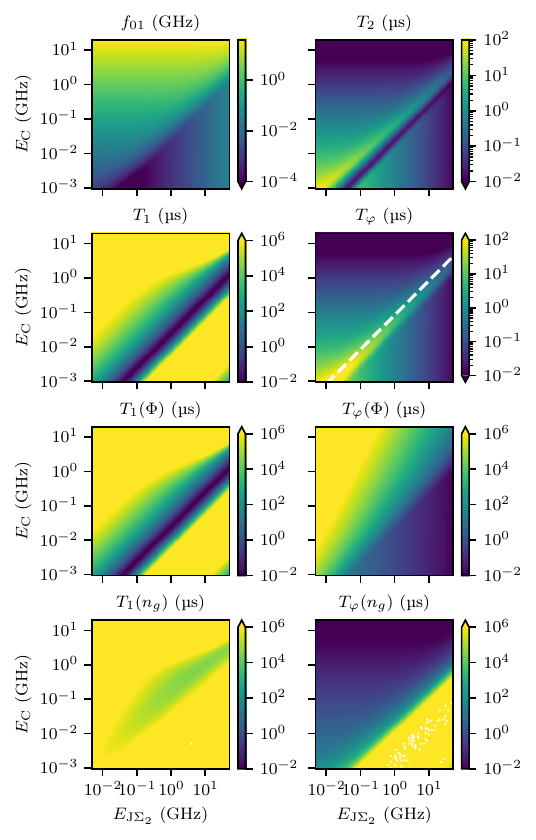}
    \caption{\textbf{Coherence characteristics times in the full parameter space.} Flux is set at $\delta \Phi = 10^{-5}$. Color scales for all $T_1$ and $T_{\varphi}$ are truncated for clarity.}
    \label{fig:Ec_Ej_opt}
\end{figure}

 \subsection*{Optimum offset flux \texorpdfstring{$\bm{\delta \Phi}$}{delta Phi} versus circuit parameters}

In Table~\ref{tab:optimized}, we showed the optimal circuit parameters $\ejsd$ and $\ec$ with respect to the longest $T_2$ at $\delta \Phi = 10^{-5}$.
We computed the circuit coherence in two sweeps.
The first one being $401 \times 401 \times 101$ points for $\ejsd$ between \SI{5}{\mega\hertz} and \SI{50}{\giga\hertz}, $\ec$ between \SI{20}{\mega\hertz} and  \SI{20}{\giga\hertz} and $\delta \Phi$ between $10^{-5}$ $\Phi_0$ and $10^{-1}$ $\Phi_0$.
The second one being $401 \times 174 \times 101$ points for $\ejsd$ between \SI{5}{\mega\hertz} and \SI{50}{\giga\hertz}, $\ec$ between \SI{1}{\mega\hertz} and  \SI{20}{\mega\hertz} and $\delta \Phi$ between $10^{-5}$ $\Phi_0$ and $10^{-1}$ $\Phi_0$.
This optimization procedure was carried out for $\ejsd/\ejsu=-0.1$.\\

\begin{figure}[ht!]
    \centering
     \includegraphics[width=\linewidth]{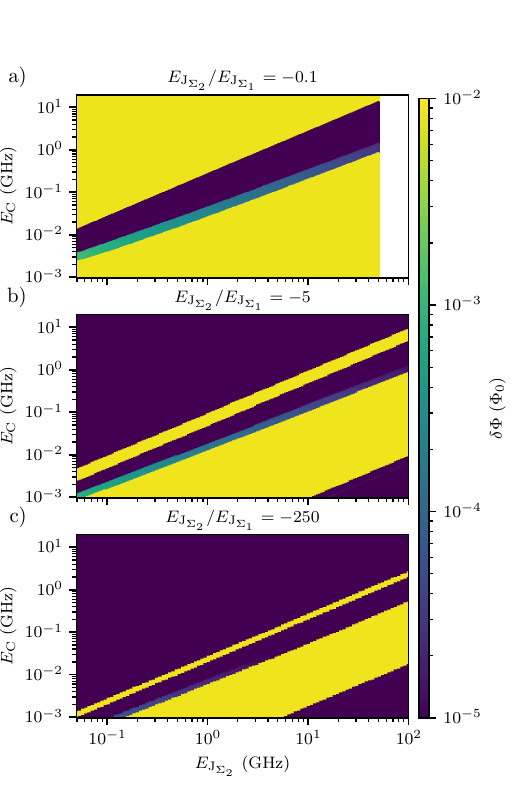}
    \caption{\textbf{Flux offset $\delta \Phi$ giving the longest $T_2$ for $\ec$ vs $\ejsu$.}
    Junctions' asymmetry is fixed to  $d=0.01$ and the charge offset $n_\mathrm{g}=0.25$.
    The ratio $\ejsd/\ejsu$ is equal to $ -0.1,- 5, -250$, from top to bottom.
    Each point corresponds to the flux offset $\delta \Phi$ found, in a range from $10^{-5}$ to $9 \times 10^{-3}$, to maximize $T_2$, see Fig.~\ref{fig:EJ2EJ1_limits}.
    }
    \label{fig:S4}
\end{figure}

We present in Fig.~\ref{fig:Ec_Ej_opt} exhaustive calculations over the $\ejsd$, $E_C$ parameter space, showing the coherence time $T_2$, the population decay time $T_1$ and the pure dephasing time $T_{\varphi}$ as well as separated contributions of charge and flux noise; for offset flux $\delta \Phi = 10^{-5}$.

On the resonance frequency plot (top left panel), we recover the two regimes of the $\cos(2\varphi)$ qubit separated by a frequency minimum, namely hybridized at low $\ejsd/E_C$ ratio (top left) and localized at large $\ejsd/E_C$ ratio (bottom right).
Accordingly, $T_{\varphi}(n_g)$ is the limiting pure dephasing time in the hybridized regime whereas $T_{\varphi}(\phi)$ is the limiting one in the localized regime.
$T_{\varphi}$ is then maximum in between the two regimes, $\ejsd/E_C \approx 12$ (See the diagonal line). We recover the global trend of increased coherence times when going to low frequency qubit, i.e. small $\ejsd$ and $E_C$.

Focusing on $T_1$, we observe a strong decrease in this region that shows limited pure dephasing, due to the contribution of 1/f flux noise at low qubit frequency. 

Combining the conclusions on $T_1$ and $T_{\varphi}$, the maximum values of $T_2$ are found for a $\ejsd/E_C$ ratio close to the ones that limit pure dephasing. At small $\ejsd \times E_C$, for which $T_1$ and $T_{\varphi}$ reach comparable values, the optimal $\ejsd/E_C$ ratio is approximately twice smaller than the one minimizing $f_{01}$, suggesting that the optimal regime is weakly hybridized between the two wells.\\

As a side note, setting the minimum achievable offset flux $\delta \Phi$ to $10^{-4}$ (instead of $10^{-5}$) moves the frontier between regimes upwards (not represented). The value of the optimal $\ejsd/E_C$ ratio shows in that case an increase of typically $50\%$, whereas the maximum $T_2$ value drops by a factor about 2, showing that the ability to control the magnetic flux value very close to $\delta \Phi = 0$ is a key experimental challenge for interference-based $\cos(2\varphi)$ qubits.

In the Fig.~\ref{fig:EJ2EJ1_limits} of the main text, we show the maximal $T_2$ for varying $\ec$ and $\ejsd$ where,  for each point, the flux offset $\delta \Phi$ is chosen in a range from $10^{-5}$ to $9 \times 10^{-3}$ to maximize $T_2$.
For the sake of completeness, we show in Fig.~\ref{fig:S4} the corresponding optimum offset flux $\delta \Phi$.
We can observe that the optimal flux for maximizing $T_2$ is almost always in one of the boundaries($10^{-5}$ or $9 \times 10^{-3}$). This suggests that, when varying the circuit parameters, there is a sharp transition between the hybridized and localized regimes as the optimal regime of operation of the $\cos(2\varphi)$ qubit.

\bibliography{bib}

@Article{Smith2020,
  author    = {Smith, W. C. and Kou, A. and Xiao, X. and Vool, U. and Devoret, M. H.},
  journal   = {npj Quantum Information},
  title     = {Superconducting circuit protected by two-Cooper-pair tunneling},
  year      = {2020},
  issn      = {2056-6387},
  month     = jan,
  number    = {1},
  volume    = {6},
  doi       = {10.1038/s41534-019-0231-2},
  publisher = {Springer Science and Business Media LLC},
}

@Article{Koch2007,
  author    = {Koch, Jens and Yu, Terri M. and Gambetta, Jay and Houck, A. A. and Schuster, D. I. and Majer, J. and Blais, Alexandre and Devoret, M. H. and Girvin, S. M. and Schoelkopf, R. J.},
  title     = {Charge-insensitive qubit design derived from the Cooper pair box},
  journal   = {Phys. Rev. A},
  year      = {2007},
  volume    = {76},
  pages     = {042319},
  month     = {Oct},
  doi       = {10.1103/PhysRevA.76.042319},
  issue     = {4},
  numpages  = {19},
  publisher = {American Physical Society},
  url       = {http://link.aps.org/doi/10.1103/PhysRevA.76.042319},
}

@article{Susskind1964,
  title = {Quantum mechanical phase and time operator},
  author = {Susskind, Leonard and Glogower, Jonathan},
  journal = {Physics Physique Fizika},
  volume = {1},
  issue = {1},
  pages = {49--61},
  numpages = {13},
  year = {1964},
  month = {Jul},
  publisher = {American Physical Society},
  doi = {10.1103/PhysicsPhysiqueFizika.1.49},
  url = {https://link.aps.org/doi/10.1103/PhysicsPhysiqueFizika.1.49}
}

@Article{Gladchenko2008,
  author    = {Gladchenko, Sergey and Olaya, David and Dupont-Ferrier, Eva and Douçot, Benoit and Ioffe, Lev B. and Gershenson, Michael E.},
  journal   = {Nature Physics},
  title     = {Superconducting nanocircuits for topologically protected qubits},
  year      = {2008},
  issn      = {1745-2481},
  month     = nov,
  number    = {1},
  pages     = {48--53},
  volume    = {5},
  doi       = {10.1038/nphys1151},
  publisher = {Springer Science and Business Media LLC},
}

@Misc{Tuokkola2024,
  author    = {Tuokkola, Mikko and Sunada, Yoshiki and Kivijärvi, Heidi and Albanese, Jonatan and Grönberg, Leif and Kaikkonen, Jukka-Pekka and Vesterinen, Visa and Govenius, Joonas and Möttönen, Mikko},
  title     = {Methods to achieve near-millisecond energy relaxation and dephasing times for a superconducting transmon qubit},
  year      = {2024},
  copyright = {Creative Commons Attribution 4.0 International},
  doi       = {10.48550/ARXIV.2407.18778},
  publisher = {arXiv},
}

@Article{Blatter2001,
  author    = {Blatter, Gianni and Geshkenbein, Vadim B. and Ioffe, Lev B.},
  journal   = {Physical Review B},
  title     = {Design aspects of superconducting-phase quantum bits},
  year      = {2001},
  issn      = {1095-3795},
  month     = apr,
  number    = {17},
  pages     = {174511},
  volume    = {63},
  doi       = {10.1103/physrevb.63.174511},
  publisher = {American Physical Society (APS)},
}

@Article{Doucot2002,
  author    = {Douçot, Benoit and Vidal, Julien},
  journal   = {Physical Review Letters},
  title     = {Pairing of Cooper Pairs in a Fully Frustrated Josephson-Junction Chain},
  year      = {2002},
  issn      = {1079-7114},
  month     = may,
  number    = {22},
  pages     = {227005},
  volume    = {88},
  doi       = {10.1103/physrevlett.88.227005},
  publisher = {American Physical Society (APS)},
}

@Article{Protopopov2004,
  author    = {Protopopov, Ivan V. and Feigel’man, Mikhail V.},
  journal   = {Physical Review B},
  title     = {Anomalous periodicity of supercurrent in long frustrated Josephson-junction rhombi chains},
  year      = {2004},
  issn      = {1550-235X},
  month     = nov,
  number    = {18},
  pages     = {184519},
  volume    = {70},
  doi       = {10.1103/physrevb.70.184519},
  publisher = {American Physical Society (APS)},
}

@article{Larsen2020,
  title = {Parity-Protected Superconductor-Semiconductor Qubit},
  author = {Larsen, T. W. and Gershenson, M. E. and Casparis, L. and Kringh\o{}j, A. and Pearson, N. J. and McNeil, R. P. G. and Kuemmeth, F. and Krogstrup, P. and Petersson, K. D. and Marcus, C. M.},
  journal = {Phys. Rev. Lett.},
  volume = {125},
  issue = {5},
  pages = {056801},
  numpages = {6},
  year = {2020},
  month = {Jul},
  publisher = {American Physical Society},
  doi = {10.1103/PhysRevLett.125.056801},
  url = {https://link.aps.org/doi/10.1103/PhysRevLett.125.056801}
}

@article{Leblanc2024,
  title = {From nonreciprocal to charge-4e supercurrent in Ge-based Josephson devices with tunable harmonic content},
  author = {Leblanc, Axel and Tangchingchai, Chotivut and Momtaz, Zahra Sadre and Kiyooka, Elyjah and Hartmann, Jean-Michel and Fernandez-Bada, Gonzalo Troncoso and Scher\"ubl, Zolt\'an and Brun, Boris and Schmitt, Vivien and Zihlmann, Simon and Maurand, Romain and Dumur, \'Etienne and De Franceschi, Silvano and Lefloch, Fran\ifmmode \mbox{\c{c}}\else \c{c}\fi{}ois},
  journal = {Phys. Rev. Res.},
  volume = {6},
  issue = {3},
  pages = {033281},
  numpages = {8},
  year = {2024},
  month = {Sep},
  publisher = {American Physical Society},
  doi = {10.1103/PhysRevResearch.6.033281},
  url = {https://link.aps.org/doi/10.1103/PhysRevResearch.6.033281}
}

@article{Leblanc2025,
  title = "Gate- and flux-tunable sin(2$\varphi$) Josephson element with planar-Ge junctions",
  author = "Leblanc, Axel and Tangchingchai, Chotivut and Sadre Momtaz, Zahra and Kiyooka, Elyjah and Hartmann, Jean-Michel and Gustavo, Fr{\'e}d{\'e}ric and Thomassin, Jean-Luc and Brun, Boris and Schmitt, Vivien and Zihlmann, Simon and Maurand, Romain and Dumur, {\'E}tienne and De Franceschi, Silvano and Lefloch, Fran{\c c}ois",
  journal = "Nat. Commun.",
  publisher = "Springer Science and Business Media LLC",
  volume =  16,
  number =  1,
  pages = "1010",
  month =  jan,
  year =  2025,
  copyright = "https://creativecommons.org/licenses/by-nc-nd/4.0",
}

@Article{Messelot2024a,
  author    = {Messelot, Simon and Aparicio, Nicolas and de Seze, Elie and Eyraud, Eric and Coraux, Johann and Watanabe, Kenji and Taniguchi, Takashi and Renard, Julien},
  journal   = {Physical Review Letters},
  title     = {Direct Measurement of a sin(2$\varphi$) Current Phase Relation in a Graphene Superconducting Quantum Interference Device},
  year      = {2024},
  issn      = {1079-7114},
  month     = sep,
  number    = {10},
  pages     = {106001},
  volume    = {133},
  doi       = {10.1103/physrevlett.133.106001},
  publisher = {American Physical Society (APS)},
}

@InCollection{Devoret1997,
  author    = {Michel H. Devoret},
  title     = {Course 10 Quantum fluctuations in electrical circuits},
  booktitle = {Quantum Fluctuations École d' été de Physique des Houches Session LXIII},
  publisher = {Elsevier},
  year      = {1997},
  editor    = {S. Reynaud, E. Giacobino and J. Zinn-Justin, eds},
  volume    = {63},
  series    = {Les Houches},
  pages     = {355 - 386},
}

@article{Hays2025,
    title = {Nondegenerate Noise-Resilient Superconducting Qubit},
  author = {Hays, Max and Kim, Junghyun and Oliver, William D.},
  journal = {PRX Quantum},
  volume = {6},
  issue = {4},
  pages = {040321},
  numpages = {19},
  year = {2025},
  month = {Oct},
  publisher = {American Physical Society},
  doi = {10.1103/dd96-gcb6},
  url = {https://link.aps.org/doi/10.1103/dd96-gcb6},
}

@Article{Somoroff2023,
  author    = {Somoroff, Aaron and Ficheux, Quentin and Mencia, Raymond A. and Xiong, Haonan and Kuzmin, Roman and Manucharyan, Vladimir E.},
  journal   = {Physical Review Letters},
  title     = {Millisecond Coherence in a Superconducting Qubit},
  year      = {2023},
  issn      = {1079-7114},
  month     = jun,
  number    = {26},
  pages     = {267001},
  volume    = {130},
  doi       = {10.1103/physrevlett.130.267001},
  publisher = {American Physical Society (APS)},
}

@article{rousseau2025enhancing,
  title={Enhancing dissipative cat qubit protection by squeezing},
  author={Rousseau, R{\'e}mi and Ruiz, Diego and Albertinale, Emanuele and d'Avezac, Pol and Banys, Danielius and Blandin, Ugo and Bourdaud, Nicolas and Campanaro, Giulio and Cardoso, Gil and Cottet, Nathanael and others},
  journal={arXiv preprint arXiv:2502.07892},
  year={2025}
}

@article{schrade2022protected,
  title={Protected hybrid superconducting qubit in an array of gate-tunable Josephson interferometers},
  author={Schrade, Constantin and Marcus, Charles M and Gyenis, Andr{\'a}s},
  journal={PRX Quantum},
  volume={3},
  number={3},
  pages={030303},
  year={2022},
  publisher={APS}
}

@Article{Bell2014,
  author    = {Bell, Matthew T. and Paramanandam, Joshua and Ioffe, Lev B. and Gershenson, Michael E.},
  journal   = {Physical Review Letters},
  title     = {Protected Josephson Rhombus Chains},
  year      = {2014},
  issn      = {1079-7114},
  month     = apr,
  number    = {16},
  pages     = {167001},
  volume    = {112},
  doi       = {10.1103/physrevlett.112.167001},
  publisher = {American Physical Society (APS)},
}

@article{Quintana2017,
  title = {Observation of Classical-Quantum Crossover of $1/f$ Flux Noise and Its Paramagnetic Temperature Dependence},
  author = {Quintana, C. M. and Chen, Yu and Sank, D. and Petukhov, A. G. and White, T. C. and Kafri, Dvir and Chiaro, B. and Megrant, A. and Barends, R. and Campbell, B. and Chen, Z. and Dunsworth, A. and Fowler, A. G. and Graff, R. and Jeffrey, E. and Kelly, J. and Lucero, E. and Mutus, J. Y. and Neeley, M. and Neill, C. and O'Malley, P. J. J. and Roushan, P. and Shabani, A. and Smelyanskiy, V. N. and Vainsencher, A. and Wenner, J. and Neven, H. and Martinis, John M.},
  journal = {Phys. Rev. Lett.},
  volume = {118},
  issue = {5},
  pages = {057702},
  numpages = {6},
  year = {2017},
  month = {Jan},
  publisher = {American Physical Society},
  doi = {10.1103/PhysRevLett.118.057702},
  url = {https://link.aps.org/doi/10.1103/PhysRevLett.118.057702}
}

@Article{Ithier2005,
  author    = {Ithier, G. and Collin, E. and Joyez, P. and Meeson, P. J. and Vion, D. and Esteve, D. and Chiarello, F. and Shnirman, A. and Makhlin, Y. and Schriefl, J. and Sch\"on, G.},
  title     = {Decoherence in a superconducting quantum bit circuit},
  journal   = {Phys. Rev. B},
  year      = {2005},
  volume    = {72},
  pages     = {134519},
  month     = {Oct},
  doi       = {10.1103/PhysRevB.72.134519},
  issue     = {13},
  numpages  = {22},
  publisher = {American Physical Society},
  url       = {http://link.aps.org/doi/10.1103/PhysRevB.72.134519},
}

@article{gyenis2021moving,
  title={Moving beyond the transmon: Noise-protected superconducting quantum circuits},
  author={Gyenis, Andr{\'a}s and Di Paolo, Agustin and Koch, Jens and Blais, Alexandre and Houck, Andrew A and Schuster, David I},
  journal={PRX Quantum},
  volume={2},
  number={3},
  pages={030101},
  year={2021},
  publisher={APS},
doi={10.1103/PRXQuantum.2.03010}
}

@article{gyenis2019experimental,
  title = {Experimental Realization of a Protected Superconducting Circuit Derived from the $0$--$\ensuremath{\pi}$ Qubit},
  author = {Gyenis, Andr\'as and Mundada, Pranav S. and Di Paolo, Agustin and Hazard, Thomas M. and You, Xinyuan and Schuster, David I. and Koch, Jens and Blais, Alexandre and Houck, Andrew A.},
  journal = {PRX Quantum},
  volume = {2},
  issue = {1},
  pages = {010339},
  numpages = {18},
  year = {2021},
  month = {Mar},
  publisher = {American Physical Society},
  doi = {10.1103/PRXQuantum.2.010339},
  url = {https://link.aps.org/doi/10.1103/PRXQuantum.2.010339}
}

@article{wang2024quantum,
  title = {Quantum control and noise protection of a Floquet $0\text{\ensuremath{-}}\ensuremath{\pi}$ qubit},
  author = {Wang, Zhaoyou and Safavi-Naeini, Amir H.},
  journal = {Phys. Rev. A},
  volume = {109},
  issue = {4},
  pages = {042607},
  numpages = {13},
  year = {2024},
  month = {Apr},
  publisher = {American Physical Society},
  doi = {10.1103/PhysRevA.109.042607},
  url = {https://link.aps.org/doi/10.1103/PhysRevA.109.042607}
}

@article{kalashnikov2020bifluxon,
  title = {Bifluxon: Fluxon-Parity-Protected Superconducting Qubit},
  author = {Kalashnikov, Konstantin and Hsieh, Wen Ting and Zhang, Wenyuan and Lu, Wen-Sen and Kamenov, Plamen and Di Paolo, Agustin and Blais, Alexandre and Gershenson, Michael E. and Bell, Matthew},
  journal = {PRX Quantum},
  volume = {1},
  issue = {1},
  pages = {010307},
  numpages = {15},
  year = {2020},
  month = {Sep},
  publisher = {American Physical Society},
  doi = {10.1103/PRXQuantum.1.010307},
  url = {https://link.aps.org/doi/10.1103/PRXQuantum.1.010307}
}

@article{spanton2017current,
  title={Current--phase relations of few-mode InAs nanowire Josephson junctions},
  author={Spanton, Eric M and Deng, Mingtang and Vaitiek{\.e}nas, Saulius and Krogstrup, Peter and Nyg{\aa}rd, Jesper and Marcus, Charles M and Moler, Kathryn A},
  journal={Nature Physics},
  volume={13},
  number={12},
  pages={1177--1181},
  year={2017},
  publisher={Nature Publishing Group UK London},
doi={10.1038/nphys4224}
}

@article{zhang2024large,
	title={{Large second-order Josephson effect in planar superconductor-semiconductor junctions}},
	author={P. Zhang and A. Zarassi and L. Jarjat and V. Van de Sande and M. Pendharkar and J. S. Lee and C. P. Dempsey and A. P. McFadden and S. D. Harrington and J. T. Dong and H. Wu and A. -H. Chen and M. Hocevar and C. J. Palmstrøm and S. M. Frolov},
	journal={SciPost Phys.},
	volume={16},
	pages={030},
	year={2024},
	publisher={SciPost},
	doi={10.21468/SciPostPhys.16.1.030},
	url={https://scipost.org/10.21468/SciPostPhys.16.1.030},
}

@article{endres2023current,
  title = {Current–Phase Relation of a WTe2 Josephson Junction},
  volume = {23},
  ISSN = {1530-6992},
  url = {http://dx.doi.org/10.1021/acs.nanolett.3c01416},
  DOI = {10.1021/acs.nanolett.3c01416},
  number = {10},
  journal = {Nano Letters},
  publisher = {American Chemical Society (ACS)},
  author = {Endres,  Martin and Kononov,  Artem and Arachchige,  Hasitha Suriya and Yan,  Jiaqiang and Mandrus,  David and Watanabe,  Kenji and Taniguchi,  Takashi and Sch\"{o}nenberger,  Christian},
  year = {2023},
  month = may,
  pages = {4654–4659}
}

@article{nanda2017current,
  title={Current-phase relation of ballistic graphene Josephson junctions},
  author={Nanda, Gaurav and Aguilera-Servin, Juan Luis and Rakyta, P{\'e}ter and Korm{\'a}nyos, Andor and Kleiner, Reinhold and Koelle, Dieter and Watanabe, Ken and Taniguchi, Takashi and Vandersypen, Lieven MK and Goswami, Srijit},
  journal={Nano Letters},
  volume={17},
  number={6},
  pages={3396--3401},
  year={2017},
  publisher={ACS Publications},
doi={10.1021/acs.nanolett.7b00097}
}

@article{gladchenko2009superconducting,
  title={Superconducting nanocircuits for topologically protected qubits},
  author={Gladchenko, Sergey and Olaya, David and Dupont-Ferrier, Eva and Dou{\c{c}}ot, Benoit and Ioffe, Lev B and Gershenson, Michael E},
  journal={Nature Physics},
  volume={5},
  number={1},
  pages={48--53},
  year={2009},
  publisher={Nature Publishing Group UK London}
}

@article{borzenets2016ballistic,
  title={Ballistic graphene Josephson junctions from the short to the long junction regimes},
  author={Borzenets, IV and Amet, F and Ke, CT and Draelos, AW and Wei, MT and Seredinski, A and Watanabe, K and Taniguchi, T and Bomze, Y and Yamamoto, M and others},
  journal={Physical review letters},
  volume={117},
  number={23},
  pages={237002},
  year={2016},
  publisher={APS}
}

@article{nguyen2019high,
  title={High-coherence fluxonium qubit},
  author={Nguyen, Long B and Lin, Yen-Hsiang and Somoroff, Aaron and Mencia, Raymond and Grabon, Nicholas and Manucharyan, Vladimir E},
  journal={Physical Review X},
  volume={9},
  number={4},
  pages={041041},
  year={2019},
  publisher={APS},
  doi       = {10.1103/physrevx.9.041041},
}

@article{wang2022hexagonal,
  title={Hexagonal boron nitride as a low-loss dielectric for superconducting quantum circuits and qubits},
  author={Wang, Joel IJ and Yamoah, Megan A and Li, Qing and Karamlou, Amir H and Dinh, Thao and Kannan, Bharath and Braum{\"u}ller, Jochen and Kim, David and Melville, Alexander J and Muschinske, Sarah E and others},
  journal={Nature materials},
  volume={21},
  number={4},
  pages={398--403},
  year={2022},
  publisher={Nature Publishing Group UK London},
doi = {10.1038/s41563-021-01187-w},
}

@unpublished{lambert2024qutip5quantumtoolbox,
      title={QuTiP 5: The Quantum Toolbox in Python}, 
      author={Neill Lambert and Eric Giguère and Paul Menczel and Boxi Li and Patrick Hopf and Gerardo Suárez and Marc Gali and Jake Lishman and Rushiraj Gadhvi and Rochisha Agarwal and Asier Galicia and Nathan Shammah and Paul Nation and J. R. Johansson and Shahnawaz Ahmed and Simon Cross and Alexander Pitchford and Franco Nori},
      year={2024},
      eprint={2412.04705},
      archivePrefix={arXiv},
      primaryClass={quant-ph},
      url={https://arxiv.org/abs/2412.04705}, 
}

@article{Chitta_2022,
doi = {10.1088/1367-2630/ac94f2},
url = {https://dx.doi.org/10.1088/1367-2630/ac94f2},
year = {2022},
month = {nov},
publisher = {IOP Publishing},
volume = {24},
number = {10},
pages = {103020},
author = {Chitta, Sai Pavan and Zhao, Tianpu and Huang, Ziwen and Mondragon-Shem, Ian and Koch, Jens},
title = {Computer-aided quantization and numerical analysis of superconducting circuits},
journal = {New Journal of Physics},
}

@unpublished{Liu2025,
  author        = {Liu, Shukai and Bordoloi, Arunav and Issokson, Jacob and Levy, Ido and Vavilov, Maxim G. and Shabani, Javad and Manucharyan, Vladimir},
  title         = {Strongly-anharmonic gateless gatemon qubits based on InAs/Al 2D heterostructure},
  year          = {2025},
  month         = mar,
  archiveprefix = {arXiv},
  copyright     = {Creative Commons Attribution 4.0 International},
  doi           = {10.48550/ARXIV.2503.12288},
  eprint        = {2503.12288},
  primaryclass  = {cond-mat.mes-hall},
  publisher     = {arXiv},
}

@article{bland20252d,
  title={2D transmons with lifetimes and coherence times exceeding 1 millisecond},
  author={Bland, Matthew P and Bahrami, Faranak and Martinez, Jeronimo GC and Prestegaard, Paal H and Smitham, Basil M and Joshi, Atharv and Hedrick, Elizabeth and Pakpour-Tabrizi, Alex and Kumar, Shashwat and Jindal, Apoorv and others},
  journal={arXiv preprint arXiv:2503.14798},
  doi           = {10.48550/arXiv.2503.14798},
  year={2025}
}

@article{you2019circuit,
  title = {Circuit quantization in the presence of time-dependent external flux},
  author = {You, Xinyuan and Sauls, J. A. and Koch, Jens},
  journal = {Phys. Rev. B},
  volume = {99},
  issue = {17},
  pages = {174512},
  numpages = {10},
  year = {2019},
  month = {May},
  publisher = {American Physical Society},
  doi = {10.1103/PhysRevB.99.174512},
  url = {https://link.aps.org/doi/10.1103/PhysRevB.99.174512}
}

@misc{Griesmar2025,
      title={Towards a $\cos(2\varphi)$ Josephson element using aluminum junctions with well-transmitted channels}, 
      author={J. Griesmar and H. Riechert and M. Hantute and A. Peugeot and S. Annabi and \c{C}. \"{O}. Girit and G. O. Steffensen and A. L. Yeyati and E. Arrighi and L. Bretheau and J. -D. Pillet},
      year={2025},
      eprint={2504.21494},
      archivePrefix={arXiv},
      primaryClass={cond-mat.mes-hall},
      doi={10.48550/arXiv.2504.21494}
}

@article{Earnest2018,
  title = {Realization of a $\mathrm{\ensuremath{\Lambda}}$ System with Metastable States of a Capacitively Shunted Fluxonium},
  author = {Earnest, N. and Chakram, S. and Lu, Y. and Irons, N. and Naik, R. K. and Leung, N. and Ocola, L. and Czaplewski, D. A. and Baker, B. and Lawrence, Jay and Koch, Jens and Schuster, D. I.},
  journal = {Phys. Rev. Lett.},
  volume = {120},
  issue = {15},
  pages = {150504},
  numpages = {6},
  year = {2018},
  month = {Apr},
  publisher = {American Physical Society},
  doi = {10.1103/PhysRevLett.120.150504},
  url = {https://link.aps.org/doi/10.1103/PhysRevLett.120.150504}
}

@article{vion2002manipulating,
  title={Manipulating the quantum state of an electrical circuit},
  author={Vion, Denis and Aassime, A and Cottet, Audrey and Joyez, Pl and Pothier, H and Urbina, C and Esteve, Daniel and Devoret, Michel H},
  journal={Science},
  volume={296},
  number={5569},
  pages={886--889},
  year={2002},
  publisher={American Association for the Advancement of Science},
doi={10.1126/science.1069372}
}

@article{yan2016flux,
  title={The flux qubit revisited to enhance coherence and reproducibility},
  author={Yan, Fei and Gustavsson, Simon and Kamal, Archana and Birenbaum, Jeffrey and Sears, Adam P and Hover, David and Gudmundsen, Ted J and Rosenberg, Danna and Samach, Gabriel and Weber, Steven and others},
  journal={Nature communications},
  volume={7},
  number={1},
  pages={12964},
  year={2016},
  publisher={Nature Publishing Group UK London}
}

@article{bylander2011noise,
  title={Noise spectroscopy through dynamical decoupling with a superconducting flux qubit},
  author={Bylander, Jonas and Gustavsson, Simon and Yan, Fei and Yoshihara, Fumiki and Harrabi, Khalil and Fitch, George and Cory, David G and Nakamura, Yasunobu and Tsai, Jaw-Shen and Oliver, William D},
  journal={Nature Physics},
  volume={7},
  number={7},
  pages={565--570},
  year={2011},
  publisher={Nature Publishing Group UK London}
}

@phdthesis{alvise,
  author  = "Alvise Borgognoni",
  title   = "Mediating high-order photon-photon interactions by Cooper-pair pairing.",
  school  = "Université Paris Sciences \& Lettres",
  year    = "2024",
}

@article{Paolo2019,
  title = {Control and coherence time enhancement of the 0–$\pi$ qubit},
  volume = {21},
  ISSN = {1367-2630},
  url = {http://dx.doi.org/10.1088/1367-2630/ab09b0},
  DOI = {10.1088/1367-2630/ab09b0},
  number = {4},
  journal = {New Journal of Physics},
  publisher = {IOP Publishing},
  author = {Paolo,  Agustin Di and Grimsmo,  Arne L and Groszkowski,  Peter and Koch,  Jens and Blais,  Alexandre},
  year = {2019},
  month = apr,
  pages = {043002}
}

@article{flowermon,
  title = {Superconducting Qubit Based on Twisted Cuprate Van der Waals Heterostructures},
  author = {Brosco, Valentina and Serpico, Giuseppe and Vinokur, Valerii and Poccia, Nicola and Vool, Uri},
  journal = {Phys. Rev. Lett.},
  volume = {132},
  issue = {1},
  pages = {017003},
  numpages = {7},
  year = {2024},
  month = {Jan},
  publisher = {American Physical Society},
  doi = {10.1103/PhysRevLett.132.017003},
  url = {https://link.aps.org/doi/10.1103/PhysRevLett.132.017003}
}

@Article{biflowermon,
  title = {Flux-tunable regimes and supersymmetry in twisted cuprate heterostructures},
  volume = {125},
  ISSN = {1077-3118},
  number = {5},
  journal = {Applied Physics Letters},
  publisher = {AIP Publishing},
  author = {Coppo,  Alessandro and Chirolli,  Luca and Poccia,  Nicola and Vool,  Uri and Brosco,  Valentina},
  year = {2024},
  month = jul,
  doi = {10.1063/5.0217614}
}

@Article{Kreikebaum2020,
  author    = {Kreikebaum, J M and O’Brien, K P and Morvan, A and Siddiqi, I},
  journal   = {Superconductor Science and Technology},
  title     = {Improving wafer-scale Josephson junction resistance variation in superconducting quantum coherent circuits},
  year      = {2020},
  issn      = {1361-6668},
  month     = apr,
  number    = {6},
  pages     = {06LT02},
  volume    = {33},
  doi       = {10.1088/1361-6668/ab8617},
  publisher = {IOP Publishing},
}

@Article{Bland2025,
  author    = {Bland, Matthew P. and Bahrami, Faranak and Martinez, Jeronimo G. C. and Prestegaard, Paal H. and Smitham, Basil M. and Joshi, Atharv and Hedrick, Elizabeth and Kumar, Shashwat and Yang, Ambrose and Pakpour-Tabrizi, Alexander C. and Jindal, Apoorv and Chang, Ray D. and Cheng, Guangming and Yao, Nan and Cava, Robert J. and de Leon, Nathalie P. and Houck, Andrew A.},
  journal   = {Nature},
  title     = {Millisecond lifetimes and coherence times in 2D transmon qubits},
  year      = {2025},
  issn      = {1476-4687},
  month     = nov,
  number    = {8089},
  pages     = {343--348},
  volume    = {647},
  doi       = {10.1038/s41586-025-09687-4},
  publisher = {Springer Science and Business Media LLC},
}

@Misc{Dane2025,
  author    = {Dane, Andrew and Balakrishnan, Karthik and Wacaser, Brent and Hung, Li-Wen and Mamin, H. J. and Rugar, Daniel and Shelby, Robert M. and Murray, Conal and Rodbell, Kenneth and Sleight, Jeffrey},
  title     = {Performance Stabilization of High-Coherence Superconducting Qubits},
  year      = {2025},
  copyright = {arXiv.org perpetual, non-exclusive license},
  doi       = {10.48550/ARXIV.2503.12514},
  publisher = {arXiv},
}

@article{zhang2021universal,
  title={Universal fast-flux control of a coherent, low-frequency qubit},
  author={Zhang, Helin and Chakram, Srivatsan and Roy, Tanay and Earnest, Nathan and Lu, Yao and Huang, Ziwen and Weiss, DK and Koch, Jens and Schuster, David I},
  journal={Physical Review X},
  volume={11},
  number={1},
  pages={011010},
  year={2021},
  publisher={APS}
}

@dataset{messelot_2025_17909031,
  author       = {Messelot, Simon and
                  Leblanc, Axel and
                  Tettekpoe, Jean-Samuel and
                  Lefloch, François and
                  Ficheux, Quentin and
                  Renard, Julien and
                  Dumur, Etienne},
  title        = {Coherence Limits in Interference-Based cos(2phi)
                   Qubits
                  },
  month        = dec,
  year         = 2025,
  publisher    = {Zenodo},
  version      = 1,
  doi          = {10.5281/zenodo.17909031},
  url          = {https://doi.org/10.5281/zenodo.17909031},
}

@article{ardati2024,
  title = {Using Bifluxon Tunneling to Protect the Fluxonium Qubit},
  author = {Ardati, Wa\"el and L\'eger, S\'ebastien and Kumar, Shelender and Suresh, Vishnu Narayanan and Nicolas, Dorian and Mori, Cyril and D'Esposito, Francesca and Vakhtel, Tereza and Buisson, Olivier and Ficheux, Quentin and Roch, Nicolas},
  journal = {Phys. Rev. X},
  volume = {14},
  issue = {4},
  pages = {041014},
  numpages = {15},
  year = {2024},
  month = {Oct},
  publisher = {American Physical Society},
  doi = {10.1103/PhysRevX.14.041014},
  url = {https://link.aps.org/doi/10.1103/PhysRevX.14.041014}
}

@article{Lin2018,
  title = {Demonstration of Protection of a Superconducting Qubit from Energy Decay},
  author = {Lin, Yen-Hsiang and Nguyen, Long B. and Grabon, Nicholas and San Miguel, Jonathan and Pankratova, Natalia and Manucharyan, Vladimir E.},
  journal = {Phys. Rev. Lett.},
  volume = {120},
  issue = {15},
  pages = {150503},
  numpages = {5},
  year = {2018},
  month = {Apr},
  publisher = {American Physical Society},
  doi = {10.1103/PhysRevLett.120.150503},
  url = {https://link.aps.org/doi/10.1103/PhysRevLett.120.150503}
}

\end{document}